\documentclass{LMCS}

\def\dOi{11(1:3)2015}
\lmcsheading%
{\dOi}
{1--40}
{}
{}
{May~\phantom{.0}4, 2011}
{Mar.~\phantom06, 2015}
{}

\usepackage{hyperref}

\usepackage{xypic}
\xyoption{curve}
\usepackage{amssymb}
\usepackage{amsmath}
\usepackage{stmaryrd}
\usepackage{color}

\newcommand{\eval}[2]{\llbracket{#1}\rrbracket_{#2}}

\newcommand{\IF}{\mathsf{IF}} \newcommand{\IC}{\mathsf{IC}}
\newcommand{\SPF}{\mathsf{SPF}} \newcommand {\JU}{J_\UU}
\newcommand{\UU}{\mathbf{U}} \newcommand{\U}{\mathsf{U}}
\newcommand{\one}{\mathrm{one}}
\newcommand{\El}{\mathsf{El}}
\newcommand{\Fam}[1]{\cat{Fam}\,#1}
\newcommand{\Id}{\mathsf{Id}}

\newcommand{\eqdf}{\mathbin{=_{\mathrm{df}}}}

\newcommand{\anon}{\_\,}

\newcommand{\alt}{^{\mathrm{alt}}}

\newcommand{\alphay}{\overline{\overline{\alpha}}}
\newcommand{\alphax}{\overline{\alpha}}
\newcommand{\lambdax}{\overline{\lambda}}
\newcommand{\lcase}[1]{{[}#1{]}}

\newcommand{\cat}[1]{\mathbf{#1}}
\newcommand{\obj}[1]{|{#1}|}
\renewcommand{\hom}[3]{{#1}\,(#2,#3)}

\newcommand{\J}{\mathbb{J}}
\newcommand{\C}{\mathbb{C}}
\newcommand{\D}{\mathbb{D}}
\newcommand{\Set}{\mathbf{Set}}
\newcommand{\Fin}{\mathbf{Fin}}
\newcommand{\X}{\mathbb{X}}
\newcommand{\cY}{\mathbb{Y}}

\newcommand{\Jfin}{J_{\mathrm{f}}}
\newcommand{\Jty}{J_{\mathrm{t}}}

\newcommand{\Kl}[1]{\mathbf{Kl}(#1)}
\newcommand{\EM}[1]{\mathbf{EM}(#1)}
\newcommand{\EMalt}[1]{\mathbf{EM}^\mathrm{alt}(#1)}

\newcommand{\Y}{\mathbf{Y}}

\newcommand{\Cat}{\mathbf{Cat}}
\newcommand{\CAT}{\cat{CAT}}

\newcommand{\Mon}[1]{\cat{Mon}(#1)}
\newcommand{\RMon}[1]{\cat{RMon}(#1)}

\newcommand{\id}{\mathrm{id}}   
\newcommand{\comp}{\circ}       % composition of maps
\newcommand{\fcomp}{\cdot}      % composition of functors
\newcommand{\fcompJ}{\fcomp^J} 

\newcommand{\idKl}{\id^T}
\newcommand{\compKl}{\mathbin{\comp^T}}

\newcommand{\Nat}{\mathbb{N}}
\newcommand{\Bool}{\mathbb{B}}

\newcommand{\op}{\mathrm{op}}

\newcommand{\inl}{\mathrm{inl}}
\newcommand{\inr}{\mathrm{inr}}

\newcommand{\Lan}{\mathrm{Lan}}

\newcommand{\Lam}{\textsf{Lam}}

\newcommand{\Ty}{\textsf{Ty}}
\newcommand{\TyLam}{\textsf{TyLam}}

\renewcommand{\lim}{\mathrm{lim}}

\newcommand{\rhoinv}{\rho^{-1}}
\newcommand{\lamxinv}{\bar{\lambda}^{-1}}
\newcommand{\alphaxinv}{\bar{\alpha}^{-1}}
\newcommand{\Tx}{T^\sharp}
\newcommand{\etax}{\eta^\sharp}
\newcommand{\mux}{\mu^\sharp}

\newcommand{\Tz}{T^\flat}
\newcommand{\etaz}{\eta^\flat}

\newcommand{\stz}{{(*^\flat)}}

\renewcommand{\Vec}{\mathsf{Vec}}
\renewcommand{\P}{\mathcal{P}}

\newcommand{\pure}{\mathsf{pure}}
\newcommand{\rcomp}{\mathbin{\lll}}

\newcommand{\m}{{\underline{m}}}
\newcommand{\n}{{\underline{n}}}

\begin{document}

\title[Monads need not be endofunctors]{Monads need not be endofunctors\rsuper*}

\author[T.\ Altenkirch]
       {Thorsten Altenkirch\rsuper a}	%required
\address{{\lsuper a}School of Computer Science, University of Nottingham, Jubilee
  Campus, Wollaton Road, Nottingham NG8 1BB, United Kingdom} %required
\email{txa@cs.nott.ac.uk}  %optional

\thanks{{\lsuper a}T.~Altenkirch was supported by the Engineering and Physical
  Sciences Research Council (EPSRC) grant no.~EP/G034109/1.}
\thanks{{\lsuper{b,c}} J.~Chapman
  and T.~Uustalu were supported by the ERDF funded Estonian CoE project EXCS,
  the Estonian Ministry of Education and Research target-financed
  themes no.\ 0322709s06 and 0140007s12, and the Estonian Science
  Foundation grants no.\ 6940, 9219 and 9475.}

\author[J.\ Chapman]{James Chapman\rsuper b}	%optional
\address{\lsuper{b,c}Institute of Cybernetics, Tallinn University of Technology,
  Akadeemia tee 21, 12618 Tallinn, Estonia} %optional
\email{\{james,tarmo\}@cs.ioc.ee}  %optional
%\thanks{}

\author[T.\ Uustalu]{Tarmo Uustalu\rsuper c}	%optional
\address{\vspace{-18 pt}} %optional
%\email{tarmo@cs.ioc.ee}  %optional
%\thanks{}
%% etc.

%% required for running head on odd and even pages, use suitable
%% abbreviations in case of long titles and many authors:

%% mandatory lists of keywords and classifications:
%\subjclass{F.3.2}
\ACMCCS{[\textbf{Software and its engineering}]: Software notations and 
tools---General programming languages---Language features---Data types 
and structures; [\textbf{Theory of computation}]: Semantics and 
reasoning---Program semantics---Categorical semantics}
\titlecomment{{\lsuper*}This article is a revised and expanded version of the FoSSaCS 2010 conference paper~\cite{thorsten.tarmo.james:fossacs10}.}
%%%%%%%%%%%%%%%%%%%%%%%%%%%%%%%%%%%%%%%%%%%%%%%%%%%%%%%%%%%%%%%%%%%%%%%%%%%

%% the abstract has to PRECEED the command \maketitle:
%% be sure not to issue the \maketitle command twice!

\begin{abstract}
  We introduce a generalization of monads, called relative monads,
  allowing for underlying functors between different categories.
  Examples include finite-dimensional vector spaces, untyped and typed
  $\lambda$-calculus syntax and indexed containers.  We show that the
  Kleisli and Eilenberg-Moore constructions carry over to relative
  monads and are related to relative adjunctions.  Under reasonable
  assumptions, relative monads are monoids in the functor category
  concerned and extend to monads, giving rise to a coreflection
  between relative monads and monads. Arrows are also an instance of
  relative monads.
\end{abstract}

\maketitle

\section{Introduction}

Monads are the most successful programming pattern arising in
functional programming. Apart from their use to model a generic notion
of effect they also serve as a convenient interface to generalized
notions of substitution. Research in the area on the border between
category theory and functional programming focusses on unveiling new
programming and reasoning constructions similar to monads, such as
% monad transformers %\cite{sheng.hudak.jones:montran}
% \cite{jaskelioff2009modular}
comonads~\cite{uustalu.vene:comoncomp},
arrows~\cite{Hug:genma} and idioms (closed
functors)~\cite{mcbride.paterson:applicative}.  Indeed, especially
when working in an expressive and total language with dependent types,
such as Agda~\cite{agda}, we can exploit monads as a way to
structure not only our programs but also their verification.

The present paper is concerned with a generalization of monads which
arises naturally in dependently typed programming, namely monad-like
entities that are not endofunctors. Consider the following example,
which arose when implementing notions related to quantum programming,
namely finite-dimensional vector
spaces~\cite{alti:qarrows,alti:qio}.
(See also
  Piponi~\cite{piponi:icfp09} for this and other interesting uses of
  vector spaces in functional programming.)

% Monads in functional programming are usually
% given as Kleisli-triples
% \begin{eqnarray*}
%   T & : & \obj{\C} \to \obj{\C}\\
%   \eta & : & \Pi_{X \in \obj{\C}}\hom{\C}{X}{T\,X} \\
%   (-)^* & : & \Pi_{X,Y \in \obj{\C}}\hom{\C}{X}{T\,Y}\to \hom{\C}{T\,X}{T\,Y}
% \end{eqnarray*}
% subject to the following laws:
% \begin{eqnarray*}
%   k^* \comp \eta  &= & k\\
%  \eta^* &= & \id\\
%   (\ell^* \comp k)^* & = & \ell^* \comp k^*
% \end{eqnarray*}
% Monads in functional programming are usually given with respect to the
% ambient category of types or sets $\Set$. An example is the list monad
% given by the triple $(T = \List,\eta = [-],k^* = \mathrm{flatten}
% \circ (\List k))$ where $\mathrm{flatten}:\Pi_{X:|\Set|}\List^2 \,X
% \to \List\,X$. Monads can also be presented with an operation $\mu :
% \Pi_{X:|\Set|} T^2 X \to T X$ instead of ${}^*$. The ${}^*$ operation
% can be derived from $\mu$ by $k^* = \mu \comp T\, k$ and indeed in the
% case of list above it is defined precisely like this. $\mu$ can also
% be derived from ${}^*$ by $\mu_X = (\id_{T\, X})^*$.

% In the case of
% relative monads on a particular functor (section 2) there is a real
% difference between the presentations as a definition with ${}^*$ does
% not automatically give rise to a $\mu$ operation.

% \txa{Shall we mention {\tt >>=}? Talk in more detail about $\mu$ and
%   relation to Kleisli triples?}

\begin{exa}\label{ex:qc}
  In quantum computing, we consider complex vector spaces, but for the
  present development any semiring $(R,0,+,1,\times)$ is
  sufficient. Finite-dimensional vector spaces (more precisely, right modules) with a given basis
can be given by:
\[
\begin{array}{l}
  \Vec \in |\Fin| \to |\Set|\\
  \Vec\, m \eqdf \Jfin\, m \to R\\
  \eta \in \Pi_{m \in |\Fin|} \Jfin\, m \to \Vec\, m\\                                                                                                        
  \eta_m\, (i \in \m) \eqdf \lambda j \in \m.\ \mathrm{if}\,
  i=j\,\mathrm{then}\,1\,\mathrm{else}\,0\\
  (-)^* \in \Pi_{m, n \in |\Fin|} (\Jfin\, m \to \Vec\, n)\to (\Vec\, m \to \Vec\, n)\\
  A^*\, x \eqdf \lambda j \in \n. \sum_{i \in \m} A\,i\,j \times x\,i 
\end{array}
\]
Here $\Fin$ is the category of finite cardinals (the skeletal version
of finite sets). The objects are natural numbers $m \in \Nat$ and the
maps between $m$ and $n$ are functions between $\m$ and $\n$ where
$\underline{m} \eqdf \{0,1,\ldots,m-1\}$. By $\Jfin \in \Fin \to \Set$
we mean the natural embedding $\Jfin\ m \eqdf \m$. The finite summation
$\sum$ is just the finite iteration of $+$ over $0$. Indeed $\eta_m$
is just the unit $m\times m$-matrix (alternatively, a function
assigning to every dimension $i \in \m$ the corresponding basis
vector) and ${A}^*\,x$ corresponds to the product of the matrix $A$
with the vector $x$, where both matrices and vectors are described as
functions.  

By the types of its data, the structure $(\Vec, \eta, (-)^*)$ looks
suspiciously like a monad, except that $\Fin$ is not $\Set$ and in the
types for $\eta$ and $(-)^*$ we have used the embedding $\Jfin$ to repair
the mismatch. It is easy to verify that the structure also satisfies
the standard monad laws, modulo the same discrepancy.

The category of finite-dimensional vector spaces with a given basis
(coordinate spaces) arises as a kind of Kleisli category. Its objects
are $m \in \Nat$, understood as finite sets of dimensions and its
morphisms are functions $\Jfin\, m \to \Vec\,n$, i.e., matrices
(describing linear transformations).

The structure cannot generally be pushed to a monad on $\Set$.
$(-)^*$ requires that we can sum over a set. Summation over general
index sets is not available, if $R$ is just a semiring. Also, in a
constructive setting, $\eta$ requires that the set has a decidable
equality, which is not the case for general sets.

Since we only require a semiring, the restrictions of the multiset and
powerset functors to $\Fin$ are instances of this construction by
using $(\Nat, 0, +, 1, \times)$ and $(\Bool, \bot,\vee,\top,\wedge) $
respectively.

We shall view $\Vec$ as a \emph{relative monad} on the embedding $\Jfin \in
\Fin\to\Set$. Other examples of relative monads include untyped and
simply typed $\lambda$-terms, the notions of indexed functors and
indexed containers as developed in \cite{alti:lics09}, and arrows.
\end{exa}

\paragraph{\bf Overview of the paper}

In Section~\ref{sec:relative-monads} we develop the notion of
relative monads on a functor $J \in \J \to \C$, showing that they
arise from relative adjunctions, and generalize the Kleisli and
Eilenberg-Moore constructions to relative monads.

Since monads on $\C$ correspond to monoids in the endofunctor category
$[\C,\C]$, a natural question is whether a relative monad on $J$
gives rise to a monoid in the category $[\J,\C]$. If $\J$ is small and
$\C$ is cocomplete (e.g., $\Set$), the left Kan extension along $J$
exists and gives rise to a left skew-monoidal structure where the unit
is $J$ and the tensor is given by $F\fcompJ G \eqdf \Lan_J\,F\fcomp
G$.\footnote{In the conference version we spoke of a `lax monoidal'
  category, but then Szlach{\'a}nyi \cite{szlachanyi} discovered the
  same structure and termed it `left skew-monoidal'.}  Relative monads
give rise to skew-monoids in this setting
(Section~\ref{sec:lax-monoids}).

Going further, we identify conditions on the functor $J \in \J \to
\C$, under which the skew-monoidal structure induced by $\Lan_J$ is
properly monoidal.  Under these well-behavedness conditions, relative
monads on $J$ are proper monoids in $[\J, \C]$. Moreover, relative
monads extend to monads via $\Lan_J$ and we get a coreflection between
the categories of relative monads on the functor $J$ and monads on the
category $\C$ (Section~\ref{sec:proper-monoids}). In the example of
vector spaces, $\Lan_J\,\Vec$ is the monad whose Kleisli category is
that of vector spaces over general sets of dimensions where a vector
over an infinite set of dimensions may only have finitely many
non-zero coordinates. However, it is worthwhile not to ignore the
non-endofunctor case, because frequently this is the structure we
actually want to use. E.g., in quantum computing we are interested in
dagger compact closed categories \cite{abramsky.coecke:quantprol}
which model finite-dimensional vector spaces.

Finally, we show that arrows are relative monads
(Section~\ref{sec:arrows}) on the Yoneda embedding. This leads to the,
maybe surprising, outcome that while arrows generalize ordinary
monads, they are actually a special case of relative monads.

\paragraph{\bf What is new?}
\label{sec:whats-new}

This paper completes the conference paper
\cite{thorsten.tarmo.james:fossacs10} with proofs, but also adds new
material. Throughout the technical part of the paper (spanning
Sections \ref{sec:relative-monads}---\ref{sec:arrows}), we
systematically speak of not just relative monads but also relative
monad morphisms, i.e., relative monads as a category, so that
restriction of monads and extension of relative monads become
functors. In Sections \ref{sec:proper-monoids}, \ref{sec:arrows} we
also accordingly treat monoid morphisms and arrow morphisms.
Discussing examples in Sections \ref{sec:relmon} and \ref{sec:reladj},
we go into more depth than in the conference paper (in particular, we
look at the EM-algebras of several examples of relative monads), but
we also consider some additional examples. In the new Sections
\ref{sec:kleisli-em-flat} and \ref{sec:kleisli-em-sharp}, we analyze
the relationship of the Kleisli and Eilenberg-Moore constructions of a
monad and its restriction and a relative monad and its extension. In
the Section~\ref{sec:alt-EM}, we present an alternative definition of
EM-algebras that is available as soon as $\Lan_J$ exists.

\paragraph{\bf Related work}
\label{sec:related-work}

The untyped $\lambda$-calculus syntax as has been identified as a
monoid in $[\Fin,\Set]$ by Fiore et al.~\cite{FPT:abssvb}. Heunen and
Jacobs~\cite{HJ:arrlma} have shown that arrows on $\C$ are actually
monoids in the category $[\C^\op \times \C, \Set]$ of endoprofunctors;
Jacobs et al.\ have proved the Freyd construction of \cite{PR:precnc}
is, in a good sense, the Kleisli construction for
arrows. Spivey~\cite{Spi:algcs} has studied a generalization of
monads, which differs from ours, but is similar in spirit and related
(see Conclusion). Berger et al. \cite{arities,grellois} have
introduced a generalization of finitary monads, called monads with
arities. Monads with arities constitute a special case of relative
monads on well-behaved functors. 

Monoidal-like categories where the unital and associativity laws are
not isomorphisms have been considered by multiple
authors. Skew-monoidal categories were introduced by Szlach\'anyi
\cite{szlachanyi} and caught then the interest of by Lack, Street,
Buckley and Garner~\cite{LS12,LS13,BGLS13}. The axioms of
skew-monoidality were also part of the axioms of the categories of
contexts of Blute et al. \cite{blute}.

\paragraph{\bf Notation}
\label{sec:background-notation}

We will be using a mixture of categorical and type-theoretic notation.
In particular we will be using $\lambda$-calculus notation for
defining functions (maps in $\Set$ or subcategories). Customarily for
both category theory and type theory, we often hide some arguments of
patterns and function applications (normally subscripted arguments,
e.g., an object a natural transformation is applied to). We write $\_$
for unique values that are easily inferable, e.g., the function on the empty domain.

We write $|\C|$ for the set of objects of $\C$ and $\hom{\C}{X}{Y}$ for the
homsets. Given categories $\C$,$\D$ we write the functor category as
$[\C,\D]$, we will also write $\C \to \D$ for the set of functors,
i.e. the objects of the functor category $[\C,\D]$. We write $\id$,
$\comp$ for the identities and composition of maps and $I$, $\fcomp$
for the identities and composition of functors.

\section{Relative monads and relative adjunctions}
\label{sec:relative-monads}

We start by defining relative monads. Then we give some examples and
show how the theory of ordinary monads carries over to the relative
case.

\subsection{Relative monads}
\label{sec:relmon}

Rather than being defined for a category $\C$ like a monad, a
relative monad is defined for a functor $J$ between two categories
$\J$ and $\C$.

\begin{defi}
A (Manes-style \cite{Man:algt}) 
\emph{relative monad} on a functor $J \in \J \to \C$ is given by
\begin{itemize}
\item an object mapping $T \in \obj{\J} \to \obj{\C}$,
\item for any $X \in \obj{\J}$, a map $\eta_X \in
  \hom{\C}{J\,X}{T\,X}$ (the \emph{unit}),
\item for any $X, Y \in \obj{\J}$ and $k \in \hom{\C}{J\,X}{T\,Y}$, a
  map $k^* \in \hom{\C}{T\,X}{T\,Y}$ (the \emph{Kleisli extension})
\end{itemize}
satisfying the conditions
\begin{itemize}
\item for any $X, Y \in \obj{\J}$, $k \in \hom{\C}{J\,X}{T\,Y}$, $k =
  k^* \comp \eta$ (the right unital law),
\item for any $X \in \obj{\J}$, $\eta_X^* = \id_{T X} \in
  \hom{\C}{T\, X}{T\, X}$ (the left unital law),
\item for any $X, Y, Z \in \obj{\J}$, $k \in \hom{\C}{J\, X}{T\, Y}$,
  $\ell \in \hom{\C}{J\, Y}{T\, Z}$, $(\ell^* \comp k)^* = \ell^*
  \comp k^*$ (the associativity law).
\end{itemize}
\end{defi}

The data and laws of a relative monad are exactly as those of a monad,
except that $\C$ has become $\J$ in some places and, to ensure
type-compatibility, some occurrences of $J$ have been
inserted. Indeed, in the laws it is only the types that have changed.

The laws imply that $T$ is functorial: $T \in \J \to \C$. Indeed, for
$X, Y \in \obj{\J}$, $f \in \hom{\J}{X}{Y}$, we can define a map $T\,
f \in \hom{\C}{T X}{T Y}$ by $T\ f \eqdf (\eta \comp J f)^*$ and this
satisfies the functor laws.  Also, $\eta$ and $(-)^*$ are natural.
% : $\eta \in \hom{[\J,\C]}{J}{T}$ and \\
% $(-)^* \in \hom{[\J^\op \times \J, \C]}{\lambda X, Y.\,
%   \hom{\C}{J\, X}{T\, Y}}{\lambda X, Y.\, \hom{\C}{T\, X}{T\,
%     Y}}$.

A definition of relative monads based on a multiplication $\mu$ rather
than a Kleisli extension $(-)^*$ is not immediately available: the
simple functor composition $T \fcomp T$ is not well-typed. In the next
section, we will show that a suitable notion of functor composition is
available under a condition.

\begin{defi}
\label{def:monad-morph}
A \emph{relative monad morphism} between two relative monads $(T, \eta,\,(-)^{*})$ and $(T',
\eta',\, (-)^{*'})$ for a particular $J$ is given by
\begin{itemize}
\item for any $X\in \obj{\J}$ a map $\sigma_X \in
\hom{\C}{T\,X}{T'\,X}$
\end{itemize} 
satisfying the conditions 
\begin{itemize}
\item for any $X\in\obj{\J},\,\sigma_X\comp{\eta}_X\,=\,{\eta'}_X$ (the unit preservation law),
\item for any $X,Y \in
\obj{\J}, k \in \hom{\C}{J\,X}{T\,Y},\,\sigma_Y\,\comp\,k^{*} =
(\sigma_Y \comp k)^{*'}\comp\,\sigma_X$ (the multiplication preservation law).
\end{itemize}
\end{defi}
The two conditions entail that $\sigma$ is natural.

It is easy to see that relative monads on a particular $J$ and
morphisms between them form a category, which we denote by
$\RMon{J}$. The identities and composition of this category are
inherited from the functor category $[\J, \C]$.

Clearly, monads on $\C$ and monad morphisms between them are a special
case of relative monads and their morphisms via $\J \eqdf \C$, $J
\eqdf I_\C$.

For general $\J$, $\C$ and $J$, we always have that $T\, X \eqdf J\,
X$ is a relative monad with $\eta_X \eqdf \id_{J X}$ and $k^* \eqdf
k$. In fact, a whole class of examples of relative monads on $J$ is
given by restricting monads on $\C$ (the relative monad $J$ arising
from restricting the monad $I_\C$).

\begin{prop} Given a functor $J \in \J \to \C$. 
\begin{enumerate}
\item A monad $(T, \eta, (-)^*)$ on $\C$ restricts to a relative monad
  $(\Tz, \etaz, (-)^\stz)$ on $J$ defined by $\Tz\, X \eqdf T\, (J\,
  X)$, $\etaz_X \eqdf \eta_{J\, X}$, $k^\stz \eqdf k^*$.

\item A monad morphism $\sigma$ between two monads $T$, $T'$ on $\C$
  restricts to a relative monad morphism $\sigma^\flat$ between the
  relative monads $T^\flat$, $T'^\flat$ on $J$ defined by
  $\sigma^\flat_X \eqdf \sigma_{J\, X}$.

\item $(-)^\flat$ is a functor from $\Mon{\C}$ to $\RMon{J}$.
\end{enumerate}
\end{prop}

The three relative monad laws and two relative monad morphism laws
follow immediately from the corresponding laws of monads and monad
morphisms.

Later we will show that, under some reasonable conditions on $J$, it
is also possible to extend relative monads to monads by a functor
$(-)^\sharp : \RMon{J} \to \Mon{\C}$. This functor is right adjoint to
$(-)^\flat$; the adjunction is a coreflection.

% \begin{thm}\label{thm:mon-restr-relmon}
%   For any $J \in \J \to \C$, a monad $(T, \eta, (-)^*)$ on $\C$
%   restricts to a relative monad $(\Tz, \etaz, (-)^\stz)$ on $J$.
% \end{thm}

% \proof 
% $\Tz\, X \eqdf T\, (J\, X)$, $\etaz_X \eqdf \eta_{J\, X}$, $k^\stz
% \eqdf k^*$. The three relative monad conditions follow immediately from
% the corresponding monad conditions.
% \qed

% This is the object map of a functor from the category of monads on
% $\C$ to the category of relative monads on $J$. The morphism map of this functor is defined as follows:

% \begin{thm}
% For any $J \in \J \to \C$, monads $T$ and $S$ on $\C$, a monad morphism $\sigma_X \in \hom{\C}{T\,X}{S\,X}$ restricts to a relative monad morphism $\sigma^\flat_X \in \hom{\C}{\Tz\,X}{S^\flat\,X}$
% \end{thm}

% \proof
% $\sigma^\flat_X\,\eqdf \sigma_{J\,X}$ The two relative monad morphism conditions follow directly from those of the monad morphism.
% \qed

As a first truly non-trivial example, we saw the relative monad of
finite-dimensional vector spaces in the introduction. Here are some
further examples.

\begin{exa}\label{ex:ulam}
  The syntax of untyped (but well-scoped) $\lambda$-calculus is a
  relative monad on $\Jfin \in \Fin \to \Set$, as is the
  finite-dimensional vector spaces relative monad, i.e., we have $\J
  \eqdf \Fin$, $\C \eqdf \Set$, $J \eqdf \Jfin$.  We view $\Fin$ as the
  category of nameless untyped contexts. The set of untyped
  $\lambda$-terms $\Lam\, \Gamma$ over a context $\Gamma$ satisfies the
  isomorphism
\[
\Lam\, \Gamma \cong \Jfin\, \Gamma + \Lam\, \Gamma \times \Lam\,
\Gamma + \Lam\, (1+\Gamma)
\]
The summands correspond to variables from the context (seen as
terms), applications, and abstractions (their bodies are terms over an
extended context). The functor $\Lam \in \Fin \to \Set$ is defined as
the carrier of the initial algebra of the functor $F \in [\Fin, \Set]
\to [\Fin, \Set]$ defined by
\[
F\, G\, \Gamma \eqdf \Jfin\, \Gamma + G\, \Gamma \times G\, \Gamma +
G\, (1+\Gamma)
\]
$\Lam$ is a relative monad. The unit $\eta \in \Jfin\, \Gamma \to
\Lam\, \Gamma$ is given by variables-as-terms and the Kleisli
extension takes a finite substitution rule $k \in \Jfin\, \Gamma \to
\Lam\, \Delta$ to the corresponding substitution function $k^* \in
\Lam\, \Gamma \to \Lam\, \Delta$.

We also introduce the relative monads $\Lam^\beta$ and
$\Lam^{\beta\eta}$ over $\Jfin$ by quotienting over $\beta$-equality
(resp.\ $\beta\eta$-equality). We observe that the monad operations
preserve the equalities, since $\beta$- and $\beta\eta$-equality are
stable under substitution.

This example was described as a relative monad (under the name
Kleisli structure) by Altenkirch and Reus~\cite{alti:csl99}.  Fiore et
al.~\cite{FPT:abssvb} described it as a monoid in a monoidal
structure on $[\Fin, \Set]$. Their account of this example is
an instance of our general description of relative monads as monoids
from Section~\ref{sec:proper-monoids}.
\end{exa}

\begin{exa}\label{ex:tlam}
  Typed $\lambda$-terms form a relative monad in a similar fashion.
  Let $\Ty$ be the set of types (over some
  base types), which we see as a discrete category. We take $\J$ to be
  $\Fin \downarrow \Ty$, which is the category whose objects are pairs
  $(\Gamma, \rho)$ where $\Gamma \in \obj{\Fin}$ and $\rho \in \Gamma
  \to \Ty$ (typed contexts) and maps from $(\Gamma, \rho)$ to
  $(\Gamma', \rho')$ are maps $f \in \hom{\Fin}{\Gamma}{\Gamma'}$ such
  that $\rho = \rho' \comp f$ (typed context maps).

  We further take $\C$ to be the functor category $[\Ty,\Set]$ and let
  $\Jty \in \Fin \downarrow \Ty \to [\Ty, \Set]$ be the natural embedding
  defined by $\Jty\, (\Gamma, \rho)\ \sigma \eqdf \{ x \in \Gamma \mid
  \rho\, x = \sigma \}$.

  Now, for $(\Gamma, \rho) \in \obj{\Fin \downarrow \Ty}$ and $\sigma
  \in \Ty$, the set of typed $\lambda$-terms $\TyLam\, (\Gamma, \rho)\,
  \sigma$ has to satisfy the isomorphism
\begin{eqnarray*}
\TyLam\, (\Gamma,\rho)\, \sigma 
& \cong &   
  \Jty\, (\Gamma, \rho)\, \sigma \\
& + & \Sigma_{\tau \in \Ty} \TyLam\, (\Gamma, \rho)\, (\tau \Rightarrow \sigma) \times \TyLam\, (\Gamma,\rho)\, \tau \\
& + & \mbox{if $\sigma$ is of the form $\tau \Rightarrow \tau'$ then~} \TyLam\, (1 + \Gamma, \left[\begin{array}{rcl} \inl\, * & \mapsto & \tau \\
                                 \inr\, x & \mapsto & \rho\, x \end{array} \right])\, \tau' 
\end{eqnarray*}
Assuming $\Id$-types the last summand can be more concisely written as:
\begin{eqnarray*}
& + & \Sigma_{\tau,\tau' \in \Ty, p \in \Id\,\sigma\,(\tau \Rightarrow \tau')}\TyLam\, (1 + \Gamma, \left[\begin{array}{rcl} \inl\, * & \mapsto & \tau \\
                                 \inr\, x & \mapsto & \rho\, x \end{array} \right])\, \tau' 
\end{eqnarray*}

The functor $\TyLam \in \Fin \downarrow \Ty \to [\Ty, \Set]$ is given
by an initial algebra. It is a monad on $\Jty$, with the unit and Kleisli
extension given by variables-as-terms and substitution, like in the
case of $\Lam$.  Fiore et al.~\cite{Fio:semane} studied $\TyLam$ as a
monoid in $[\Fin \downarrow \Ty, [\Ty, \Set]]$. As in Example
\ref{ex:ulam} we can quotient by $\beta$- or $\beta\eta$-equality and
as before we denote the corresponding relative monads as $\TyLam^\beta$
and $\TyLam^{\beta\eta}$.

Note that choosing $\J$ to be $[\Ty,\Fin]$ rather than $\Fin
\downarrow \Ty$ would have given contexts possibly supported by
infinitely many types: in every type there are finitely many
variables, but the total number of variables can be infinite.
\end{exa}

\begin{exa}\label{ex:ic}
  Morris and Altenkirch~\cite{alti:lics09} investigated a generalization
  of the notion of containers \cite{alti:cont-tcs} to a dependently
  typed setting and used it to show that strictly positive families
  can be reduced to W-types. Relative monads played a central role in
  this development.

  Let $\U\in\obj{\Set}$ together with a family $\El\in \U\to\obj{\Set}$ 
  which we view as a type theoretic universe. As an example  consider the universe
  of small sets which reflects all type theoretic constructions but $\U$ itself. 
  E.g., there is $\pi \in \Pi A \in \U.\, (\El\, A \to \U) \to \U$ such that 
  $\El\,(\pi\,A\,B)$ is isomorphic to $\Pi a \in \El\,A.\, \El\,(B\,a)$.
  And similarly for the other type formers.

  Such a universe induces a category $\UU$ with $|\UU| \eqdf \U$ and
  $\hom{\UU}{A}{B} \eqdf \El\,A \to \El\,B$. The functor $\JU \in
  \UU \to \Set$ is given by $\JU\,A \eqdf \El\,A$ on objects and the
  identity on maps.\footnote{In \cite{alti:cont-tcs} we actually
    used $\Set$ and $\Set_1$ instead of $\U$ and $\Set$, and $\El$ is
    usually implicit in type theory.}  We assume that $\UU$ is locally
  cartesian closed, which corresponds to the assumption that $\U$
  reflects $\Pi$, $\Sigma$ and equality types.

  While ordinary containers represent endofunctors on $\UU$, indexed
  containers represent functors from the category of families
  $\Fam{A}$ to $\UU$ for $A\in\U$. $\Fam{A}$ has as objects families
  $\El\,A \to \U$ and as morphisms between $F,G \in \El\,A \to
  \U$ families of functions $\Pi
  a\in\El\,A.\, \El\,(F\,a ) \to \El\,(G\,a)$.
  Indeed, $\Fam{A}$ is equivalent to the slice category
  $\UU/A$. For $A \in \U$, we define the set of indexed functors by  $\IF\,A \eqdf \Fam{A} \to\UU$  % with $\IF\,A
%   \in \UU \to \Set$ 
  and indeed $\IF$ gives rise to a relative monad
  on $\JU$: The unit $ \eta_A \in \JU\,A \to \IF\,A$ is defined by
  $\eta_A\,x \eqdf \lambda f.\, f\,x$ and the Kleisli extension $k^*
  \in \IF\,A \to \IF\,B$ of $k \in\JU\,A \to \IF\,B$ is defined by
  $k^*\,G\,f \eqdf G\,(\lambda x.k\,x\,f)$.  The definitions clearly
  resemble the continuation monad apart from the size issue.

  The main result of \cite{alti:lics09} was that strictly positive
  families ($\SPF$) can be interpreted as indexed functors via
  indexed containers ($\IC$). Just as $\IF$, both $\SPF$ and $\IC$ are
  relative monads on $\JU$ and the interpretations preserve this
  structure, i.e., are relative monad maps. The relative monads
  model the fact that all these notions are closed under substitution
  and that this is preserved by the constructions done in the paper.
 \end{exa}

\begin{exa}
  Sam Staton suggested to us this example of a ``relative monad'' that
  naturally arises at a different level---a relative pseudo-monad, in
  fact. Let $\J \eqdf \Cat$ (the category of small categories), $\C
  \eqdf \CAT$ (the category of locally small categories) and let $J \in
  \Cat \to \CAT$ be the inclusion.

  Define $T \in \Cat \to \CAT$ by $T\, \X \eqdf [\X^\op, \Set]$, i.e.,
  $T$ sends a given small category to the corresponding presheaf
  category, which is locally small. 

  We can define $\eta_\X \in \X \to [\X^\op, \Set]$ to be $\Y_\X$ (the
  Yoneda embedding for $\X$). And, for $K \in \X \to [\cY^\op,
  \Set]$, we can set $K^* \in [\X^\op, \Set] \to [\cY^\op, \Set]$ to be 
  $\Lan_{\Y_\X}\, K$.

  Note that the laws do not hold on the nose now, but only up to
  coherent isomorphism.
\end{exa}

\begin{exa}
\label{ex:state-cont}
We already know that, for any monad $T$ on $\C$, $T^\flat = T \cdot J
\in \J \to \C$ is a relative monad on $J$.
Interesting special cases of this basic observation arise already when
$T$ is the identity functor on $\C$. E.g., we can take $\C \eqdf
\Set$, $\J \eqdf \Set$, $J\, X \eqdf X \times S$, $T^\flat\, X \eqdf J\, X$
for some fixed $S \in |\Set|$.  Or we can take $\C \eqdf \Set$, $\J
\eqdf \Set^\op$, $J X \eqdf R^X$ for some fixed $R \in |\Set|$. As we
will see below, these constructions behave in some aspects like the
state and continuations monads.
\end{exa}

\subsection{Relative adjunctions}
\label{sec:reladj}

As ordinary monads are intimately related to adjunctions, relative
monads are related to a corresponding generalization of
adjunctions. Similarly to the situation with relative monads, not
every definition format of adjunctions is available for relative
adjunctions, if we make no assumptions about $J$: definitions
involving a counit are not possible. The following is one of the
possible definitions.

\begin{defi}
A \emph{relative adjunction} between $J \in \J \to
\C$ and $\D$ is given by two functors $L  \in \J \to \D$ and $R \in \D \to
\C$ and a natural isomorphism
$ %\[
\phi_{X,Y} \in \hom{\C}{J\, X}{R\, Y} \cong \hom{\D}{L\, X}{Y}
$. %\]
\end{defi}

As expected, ordinary adjunctions are a special case of relative
adjunctions with $\J \eqdf \C$, $J \eqdf I$. 
Just like any adjunction defines a monad, relative adjunctions define
relative monads.

\begin{thm}
  Any relative adjunction $(L, R, \phi)$ between a functor $J \in \J
  \to \C$ and category $\D$ gives rise to a relative monad $(T, \eta,
  (-)^*)$ via $T\, X \eqdf R\, (L\, X)$, $\eta_X \eqdf \phi^{-1}\,
  (\id_{L\, X})$ and $k^* \eqdf R\, (\phi\, k)$.

\[
\xymatrix@R=1pc{&\mathbb{D}\ar@/^/[rd]_R\\
          \mathbb{J}\ar@/^/[ru]_L \ar[rr]^J \ar@/^10ex/[rr]^T&&\mathbb{C}}
\]
\end{thm}

\proof We have to check the relative monad laws. The right unital law
holds since $k = \phi^{-1}\, (\phi\, k) = \phi^{-1}\, (\phi\, k \comp
\id_{L\, X}) = R\, (\phi\, k) \comp \phi^{-1} \id_{L\, X} = k^* \comp
\eta_X$ by $\phi^{-1}$ being a left inverse of $\phi$ and $\phi^{-1}$
being natural.

The left unital law is verified by $(\eta_X)^* = R\, (\phi\,
(\phi^{-1}\, \id_{L\,X})) = R\, \id_{L\, X} = \id_{R\, (L\,X)} = \id_{T\, X}$ by
$\phi^{-1}$ being a right inverse of $\phi$ .

For the associative law we calculate $(\ell^* \comp k)^* = R\, (\phi\,
(R (\phi\, \ell) \circ k)) = R\, (\phi\, \ell \comp \phi\, k) = R
(\phi\, \ell) \circ R (\phi\,k) = \ell^* \comp k^*$ by $\phi$ being
natural.  \qed

If a relative monad $T$ on $J$ is related to a relative adjunction
$(L, R, \phi)$ between $J$ and some category $\D$ in the above way, we
call the relative adjunction a \emph{splitting} of the relative monad
via $\D$.

\subsection{Kleisli and Eilenberg-Moore constructions}

We know that a monad splits into an adjunction in two
canonical ways: the Kleisli and Eilenberg-Moore
constructions. Moreover, the splittings form a category where the
Kleisli and EM splittings are the initial and terminal objects. We
shall now establish that the same holds in the relative situation.

The Kleisli category $\Kl{T}$ of a relative monad $T$ has as objects
the objects of $\J$ and as maps between objects $X$ and $Y$ of $\J$ the maps between objects $J\
X$, $T\, Y$ of $\C$: $\obj{\Kl{T}} \eqdf \obj{\J}$ and
$\hom{\Kl{T}}{X}{Y} \eqdf \hom{\C}{J\, X}{T\, Y}$. The identity and
composition (we denote them by $\idKl$, $\compKl$) are defined by
$\idKl_X \eqdf \eta_X$ and $\ell \compKl k \eqdf \ell^* \comp k$.

The Kleisli relative adjunction between $\J$ and $\Kl{T}$ is defined by
%\begin{eqnarray*}
$L\, X  \eqdf  X$, 
$L\, f  \eqdf  \eta \comp J f$ (note that $L$ is identity-on-objects), 
$R\, X  \eqdf  T\, X$, 
$R\, k  \eqdf  k^*$
and $\phi$ is identity.
%$\phi_{X,Y}\, k  \eqdf  k$,
%$\phi^{-1}_{X, Y}\, k  \eqdf  k$ 
%\end{eqnarray*}
%
%Note especially that $L$ is identity-on-objects and the isomorphism
%$\phi$ is nothing but identity. 
This relative adjunction is a splitting. Indeed, we have $R\, (L\, X)
= T\, X$, $R\, (L\, f) = (\eta \comp J\, f)^* = T\, f$, $\eta_X =
\idKl_X = \phi^{-1}\, (\idKl_{X}) = \phi^{-1}\, (\idKl_{L\, X})$ and
$k^* = R\, k = R\, (\phi\, k)$.

The Eilenberg-Moore (EM) category $\EM{T}$ is given by EM-algebras and
EM-algebra maps of the relative monad $T$. Since the usual definition
of an EM-algebra refers to $\mu$, which is not immediately available,
we generalize a version based on $(-)^*$. For ordinary monads this is
equivalent to the standard definition.

\begin{defi}\label{def:EM-algebra}
  An EM-\emph{algebra} of a relative monad $T$ on $J \in \J \to \C$ is
  given by an object $X \in \obj{\C}$ (the \emph{carrier}) and, for
  any $Z \in \obj{\J}$ and $f \in \hom{\C}{J\, Z}{X}$, a map $\chi\, f
  \in \hom{\C}{T\, Z}{X}$ (the \emph{structure}), satisfying the
  conditions
\begin{itemize}
\item for any $Z \in \obj{\J}$, $f \in \hom{\C}{J\, Z}{X}$, $f = \chi\, f
  \comp \eta$,
\item for any $Z, W \in \obj{\J}$, $k \in \hom{\C}{J\, Z}{T\, W}$, $f
  \in \hom{\C}{J\, W}{X}$, $\chi\, (\chi\, f \comp k) = \chi\, f \comp k^*$.
\end{itemize}
These conditions ensure, among other things, that $\chi$ is natural.

An EM-\emph{algebra map} from $(X, \chi)$ to $(Y, \upsilon)$ is a map $h \in
\hom{\C}{X}{Y}$ satisfying
\begin{itemize}
\item for any $Z \in \obj{\J}$, $f \in \hom{\C}{J\, Z}{X}$, $h \comp
  \chi\, f = \upsilon\, (h \comp f)$.
\end{itemize}
\end{defi}

The identity and composition maps of $\EM{T}$ are inherited from $\C$.

The Eilenberg-Moore relative adjunction between $J$ and $\EM{T}$ is
defined by
%\begin{eqnarray*}
$L\, X  \eqdf  (T\, X, \lambda k.\, k^*)$,
$L\, f  \eqdf  T\,  f$,
$R\, (X, \chi)  \eqdf  X$,
$R\, h  \eqdf  h$, 
$\phi_{X, (Y,\upsilon)}\, f \eqdf \upsilon\, f$ and 
$\phi^{-1}_{X, (Y,\upsilon)}\, h \eqdf h \comp \eta_X$.
%\end{eqnarray*}
%
%Here, $R$ is identity-on-maps and $L$ on maps is $T$. 
This is also a splitting. 

% Indeed: $R\, (L\, X) = T\, X$, $R\, (L\, f)
% = T\, f$, $\eta_X = \id_{T\, X} \comp \eta_X = \phi^{-1} \id_{(T\,X,
%  \lambda k.\, k^*)} = \phi^{-1} \id_{L\, X}$ and ...

%\[
%\begin{array}{llcl}
%{F^*}^T&&:&\mathbb{J} \rightarrow \mathbb{C}^T\\
%{F^*}^T&A&=&(T A, \lambda_X f : J X \rightarrow T A .\,f^*)\\
%{F^*}^T&f&=& T f\\\\
%{U^*}^T&&:&\mathbb{C}^T\rightarrow \mathbb{C}\\
%{U^*}^T&(A, a)&=&A\\
%{U^*}^T&f&=&f
%\end{array}
%\]

%\todo{How does this fit in?} It doesn't. It just repeats the theorem below.
%The splittings form a category where an object is a relative adjunctions for a fixed
%$\J$, $\C$, and $J$ which we represent as the tuple $(\D,L : \J \to
%\D, R : \D \to \C, \phi : \hom{\C}{J\, X}{R\, Y} \cong \hom{\D}{L\,
%  X}{Y})$ where $T = R \circ L$, $\eta_X = \phi^{-1}\, (\id_{L\, X})$,
%and $k^* = R\, (\phi\, k)$.  Given two objects in this category
%$(\D,L,R,\phi)$ and $(\D',L',R',\phi')$ a morphism is a functor $V :
%\D \to \D'$ such that $L'\,=\,V\circ L$, $R\,=\,R' \circ V$, and
%$\phi'_{X,V\,Y}\,=\,V\,\phi_{X, Y}$.

\begin{thm}
  The splittings of a relative monad $T$ on $J \in \J \to \C$ form a
  category. An object is given by a category $\D$ and an adjunction
  $(L, R, \phi)$ splitting $T$ via $\D$. A splitting morphism between $(\D,
  L, R, \phi)$ and $(\D',L',R',\phi')$ is a functor $V \in
  \D \to \D'$ such that $V \fcomp L = L'$, $R = R' \fcomp V$, and
  $V\, (\phi_{X, Y}\, k) = \phi'_{X,V\,Y}\, k$.  The Kleisli construction is the
  initial and the Eilenberg-Moore construction the terminal splitting.
\end{thm}

\proof 
To show that the Kleisli splitting is initial we show that the
following is a unique morphism from the Kleisli splitting
$(\Kl{T},L_T,R_T,\phi_T)$ to a given other splitting $(\D,L, R,
\phi)$.  We define:
\[
\begin{array}{lcl}
V & \in &\Kl{T} \to\D\\
V \,X&=&L\,X\\
V \,k&=&\phi_{X,L Y}\,k
\end{array}
\]
The functoriality of $V$ is verified by $V\, \eta = \phi\, \eta =
\id$, $V\, (\ell^* \comp k) = \phi \, (R\, (\phi\, \ell) \comp k) =
\phi\, \ell \comp \phi\, k$. 

The splitting morphism conditions are verified by $V\, (L_T\, X) = L\,
X$, $V\, (L_T\, f) = \phi\, (\eta \comp J\, f) = \phi\, (\phi^{-1}\,
\id \comp J\, f) = \phi\, (\phi^{-1}\, (L\, f)) = L\, f$, $R_T\, X =
T\, X = R\, (L\, X) = R\, (V\, X)$, $R_T\, k = k^* = R\, (\phi\, k) =
R\, (V\, k)$, $V\, (\phi_T\, k) = \phi\, k$.

Uniqueness is established as follows. Any morphism $V'$ between the
two splittings must satisfy $V'\, X = V'\, (L_T\, X) = L\, X = V\, X$,
$V'\, k = V'\, ((\phi_T)_{X,Y}\, k) = V'\, \phi_{X,V\, Y}\, k = \phi_{X,L\,
  Y}\, k = V\, k$. 

For finality of the EM splitting we prove that the following is a
unique morphism between a given splitting $(\D, L,R, \phi)$ and the
EM splitting $(\EM{T},L^T,R^T,\phi^T)$. We set
\[
\begin{array}{lcl}
V & \in &\D \to \EM{T}\\
V \,X& \eqdf & (R\,X, \lambda k.\, R\, (\phi\, k)) \\
%\lambda X:\obj{\J},k:\hom{\C}{J\,Z}{R X}.\;R\,(\phi\,k))\\
V \,f& \eqdf &R\,f\\
\end{array}
\]
That $V\, X$ is an EM-algebra is seen by checking that $R (\phi\, k)
\comp \eta = k^* \comp \eta = k$, $R\, (\phi\, (R\, (\phi\, \ell)
\comp k)) = (\ell^* \comp k)^* = \ell^* \comp k^* = R\, (\phi\, \ell)
\comp k^*$ using that $(\D, L, R, \phi)$ is a splitting. The
functoriality of $V$ follows immediately from the functoriality of
$R$.

The conditions of a splitting morphisms are verified by $V\, (L\, X) =
(R\, (L\, X), \lambda k.\, R\, (\phi\, k)) = (T\, X, \lambda k.\, k^*) =
L^T\, X$, $V\, (L\, f) = R\, (L\, f) = T\, f = L^T\, f$, 
$R\, X = R^T\, (R\, X, \lambda k.\ R\, (\phi\, k)) = R^T\, (V\, X)$, 
$R\, f =
R^T\, (V\, f)$, $V\, (\phi\, k) = R\, (\phi\, k) = k^\star = \phi^T\,
k$.

For uniqueness we observe that any splitting $V'$ must satisfy $V' X =
(V'_0\, X, V'_1\, X) = (V'_0\, X, \lambda k.\, \phi^T\, k) = (R^T\,
(V'\, X), \lambda k.\, V'\, (\phi\, k)) = (R\, X, \lambda k.\, R\,
(\phi\, k)) = V\, X$.
\qed

The Kleisli and Eilenberg-Moore categories of our examples correspond
to well known concepts.

\begin{exa}  
  The Kleisli category of $\Vec$ has as objects the objects of $\Fin$
  understood as finite sets of dimensions.  The maps are maps $\Jfin\,
  m \to \Vec\, n$, i.e., $m \times n$-matrices (describing linear
  transformations). The identities are the unit $m \times m$-matrices,
  the composition is multiplication of matrices. It is the category of
  finite-dimensional coordinate spaces and linear transformations.
\end{exa}

\begin{exa}
  The Kleisli category $\Kl{\Lam}$  of the
  relative monad for untyped $\lambda$-terms (Example \ref{ex:ulam}) has as objects
  the objects of $\Fin$ understood as untyped contexts. The maps are
  maps $\Jfin\, m \to \Lam\, n$, i.e., substitution rules
  (assignments of terms over $n$ to the variables in $m$). The
  identities are the trivial substitution rules. The composition is
  composition of substitution rules.
\end{exa}
% changed "substitution rules" to substitutions to be consistent with the rest of the paper (and general use). txa

\begin{exa}
  The Kleisli category $\Kl{\TyLam}$ of the relative monad for typed
  $\lambda$-terms (Example \ref{ex:tlam}) has a very similar
  structure. Its objects are typed contexts, i.e., objects of $\Fin
  \downarrow \Ty$, and its morphisms are type-preserving
  substitution rules. Indeed, the Kleisli category of $\TyLam^{\beta\eta}$
  is equivalent to the free cartesian closed category on the set of base types (if we also include finite products into the type language and amend the term language accordingly). %CHECK - something's wrong!!
\end{exa}

\begin{exa}
  The Kleisli categories of the two relative monads considered in
  Example~\ref{ex:state-cont} are isomorphic to those of the ordinary
  state and continuation monads.

  For $J X \eqdf X \times S$, $T X \eqdf X \times S$, $T$ is a
  relative monad on $J$ and the maps of its Kleisli category are maps
  $X \times S \to Y \times S$. But the ordinary state monad $T'$ given
  by $T' X \eqdf (X \times S)^S$ has maps $X \to (Y \times S)^S$ as
  the maps of its Kleisli category. Clearly, the two categories are
  isomorphic.
  
  We also get such an isomorphism for the Kleisli categories of the
  relative monad $T$ given by $T X \eqdf R^X$ on $J$ given by $J X \eqdf R^X$ and the ordinary
  continuations monad $T'$ given by $T' X \eqdf R^{R^X}$.
\end{exa}

\begin{exa}
\label{exa:vec-alg}
A vector space (right module) over a semiring $(R,0,+,1,\times)$ is given by a commutative
monoid $(M,\vec{0},\oplus)$ and an operation $\cdot \in M \times R
\rightarrow M$. It is isomorphic to a relative EM-algebra for the
relative monad $\Vec$ over the semiring $R$. The carrier of the
algebra is defined to be $M$ and the structure map $m \in \Pi_{n \in
  |\Fin|} (\Jfin\, n \to M)\to (\Jfin\, n \to R)\to M$ is given by
lifting the operation $\cdot$ straightforwardly to an operation on
vectors: $m_n\,f\,g\eqdf \bigoplus_{i \in \n} f\,i\,\cdot\,g\,i$. Going the
other way, given an algebra with carrier $M$ and structure map $m$,
$\vec{0} \eqdf m_0\,\_\,\_$, $a \oplus a' \eqdf m_2\,(\lambda i.\,
\mathrm{if}\,i = 0\, \mathrm{then}\, a\, \mathrm{else}\, a')
(\lambda i .\, 1)$  and $a \cdot r \eqdf m_1\,(\lambda i .\,
a)\,(\lambda i.\, r)$.
\end{exa}

\begin{exa}
\label{exa:wslam-alg}
The objects of $\EM{\Lam^\beta}$ correspond to $\lambda$-models,
e.g., as given in definition 11.3 in \cite[p. 112]{hindley.seldin}.
An EM-algebra is given by a set $D$ and for any $n \in |\Fin|=\Nat$
a function
\[ \delta \in (\Jfin\, n \to D) \to (\Lam^\beta\,n \to D) \]
subject to the two conditions stated in definition
\ref{def:EM-algebra}. This gives rise to a $\lambda$-model with
carrier $D$, the applicative structure can be obtained from $\delta$ and
$\delta$ also gives rise to the evaluation function simply by 
$\eval{t}{\rho} = \delta\,\rho\,t$. The conditions for a $\lambda$-model
follow from the conditions of the EM-algebra. The evaluation function
in \cite{hindley.seldin} is not scoped, but it can be seen that the explicit
indexing corresponds to the variable condition (e). On the other hand we
can obtain an EM-algebra from a $\lambda$-model in the sense of 
\cite{hindley.seldin}. We can also show that objects of
$\EM{\Lam^{\beta\eta}}$ correspond to extensional $\lambda$-models.
\end{exa}

\begin{exa}\label{exa:wtlam-alg}
  In a similar way, the objects of $\EM{\TyLam^\beta}$ correspond to
  type frames as given in \cite[p.\ 53]{gunter}. The carrier of an
  EM-algebra corresponds to the interpretation of types given by a
  preframe ${\mathcal A}^{\textrm{type}}$, while the structure
    corresponds to the interpretation of terms ${\mathcal
      A}^{\textrm{term}}$. The objects of $\EM{\TyLam^{\beta\eta}}$
      correspond to extensional type frames.
\end{exa}

\begin{exa} An algebra of the first of the two relative monads $T$ of
  Example~\ref{ex:state-cont} is a pair $(X, \chi \in \int_Z ((Z
  \times S \to X) \to (Z \times S \to X))$. As 
$ 
\int_Z ((Z \times S \to X) \to (Z \times S \to X)) 
\cong (\int^Z (Z \times S \to X) \times Z \times S) \to X 
\cong (\int^Z (Z \to X^S) \times (Z \times S)) \to X 
\cong X^S \times S \to X 
$, 
this is the same as to give a pair $(X, x \in X^S \times S \to X)$.

The algebras of the state monad $T'\, X \eqdf (X \times S)^S$ are pairs
$(X, x \in (X\times S)^S \to X)$. We can see that the two EM
categories are not equivalent.
\end{exa}

\subsection{Kleisli and Eilenberg-Moore constructions and restriction}
\label{sec:kleisli-em-flat}

What is the relationship between the Kleisli and Eilenberg-Moore
constructions of some given monad $T$ on $\C$ and the relative monad
$T^\flat$ on $J$?

There is a functor $D \in \Kl{T^\flat} \to \Kl{T}$ defined as follows:
\begin{itemize}
\item for any $X \in |\J|$, $D\, X \eqdf J\, X$,
\item for any $X, Y \in |\J|$, $k \in \C(J\, X, T\, (J\, Y))$,  $D\, k \eqdf k$
\end{itemize}
No assumptions are needed to prove that $D$ preserves the identities
and composition of $\Kl{T^\flat}$.

Let $L, R$ be the Kleisli adjunction of $T$, which is given by $L\, X
\eqdf X, L\, f \eqdf \eta \comp f, R\, X \eqdf T\, X, R\, k \eqdf k^*$.

The relative Kleisli adjunction of $T^\flat$ is given by
$L'\, X \eqdf X, L'\, f \eqdf \eta^\flat \comp J\, f = \eta \comp J\, f,
R'\, X \eqdf T^\flat X = T\, (J\, X), R'\ k \eqdf k^{(*^\flat)} = k^*$.

We have $D \cdot L' = L \cdot J$ and $R' = R \cdot D$.  
Moreover, the category $\Kl{T}$ together with the functors $L \cdot J$
and $R$ gives a splitting of $T^\flat$: we have $R \cdot (L \cdot J)
= T \cdot J$ and $L \cdot J$ is relative left adjoint to $R$.

In general we can define no functor in the opposite direction $\Kl{T}
\to \Kl{T^\flat}$, for the simple reason that this would require some
canonical functor $\C \to \J$ and we have none given.

There is also a functor $E \in \EM{T} \to \EM{T^\flat}$ defined by
\begin{itemize}
\item for any $(X, x) \in |\EM{T}|$, i.e., $X \in |\C|$, $x \in \C(TX,
  X)$ meeting the EM-algebra conditions, $E(X,x) \eqdf (X, \chi)$
  where for $Z \in |\J|$, $f \in \C(JZ,X)$, $\chi_Z\, f \eqdf x \comp
  T\, f \in \C(T(JZ), X)$; \\
  $E(X,x)$ is a relative EM-algebra of $T^\flat$ under no assumptions;
\item for any $h \in \EM{T}((X,x),(Y,y))$, i.e., $h \in \C(X,Y)$ meeting the EM-algebra morphism conditions, $E\, h \eqdf h$; \\
  $E\, h$ satisfies the relative EM-algebra conditions.
\end{itemize}
It is trivial that $E$ preserves the identities and composition of
$\EM{T}$.

Let $F$, $U$ be the EM adjunction of $T$, which is given by $F\, X
\eqdf (T\,X, \mu_X)$, $F\, f = T\, f$, $U\, (X, x) \eqdf X$, $U\, h \eqdf
h$. 

The relative EM adjunction of $T^\flat$ is given by 
$F'\, X \eqdf (T\, (J\, X), (-)^*)$, $F'\, f \eqdf T\, (J\, f)$, 
$U'\, (X, \chi) \eqdf X$, $U'\, h \eqdf h$.

We have $F' = E \fcomp (F \fcomp J)$ and $U' \fcomp E = U$. 
Furthermore, the category $\EM{T}$ together with the functors $F
\fcomp J$ and $U$ gives a splitting for $T^\flat$: we have $U \fcomp
(F \fcomp J) = T \fcomp J$ and $F \fcomp J$ is relative left
adjoint to $U$.

In general, we cannot construct a functor $\EM{T^\flat} \to \EM{T}$.

This situation is illustrated on the following diagram.
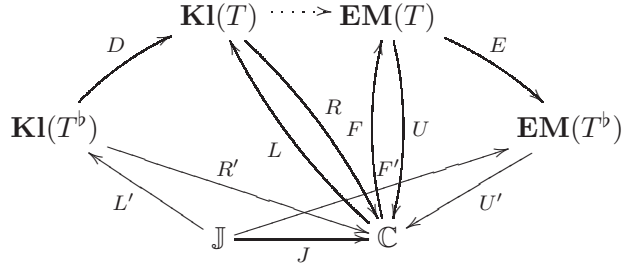
\[
\xymatrix{
& \Kl{T} \ar@/^0.2pc/@{.>}[r] \ar@/^0.5pc/[ddr]^{R}
  & \EM{T} \ar@/^0.5pc/[dr]^{E} \ar@/^0.5pc/[dd]^{U}
    &  \\
\Kl{T^\flat} \ar@/^0.5pc/[ur]^{D} \ar[drr]^{R'}
&  
  & 
    & \EM{T^\flat} \ar[dl]^{U'}\\   
& \J \ar[r]_{J} \ar[ul]^{L'} \ar[urr]^{F'}
  & \C \ar@/^0.5pc/[uul]^{L} \ar@/^0.5pc/[uu]^{F}
    & 
}
\]

% =======================================================================
% =======================================================================
% =======================================================================

\section{Relative monads as skew-monoids in a skew-monoidal category}
\label{sec:lax-monoids}

A monad on $\C$ is the same as a monoid in the endofunctor category
$[\C,\C]$. This category has a monoidal structure given by the identity
functor $I$ and composition of functors $\fcomp$, which are strictly
unital and associative. A monad can be specified by an object
$T \in \obj{[\C,\C]}$ and maps $\eta \in \hom{[\C,\C]}{I}{T}$ and $\mu
\in \hom{[\C,\C]}{T \fcomp T}{T}$ satisfying the laws of a monoid in
the strict monoidal category $([\C,\C],I, \fcomp)$.

Can we similarly define a relative monad on $J \in \J \to \C$ as a
monoid in the functor category $[\J,\C]$? This requires a monoidal
structure on $[\J,\C]$, ideally similar to that on $[\C,\C]$.  The
functor $J$ is a good candidate for the unit, but the tensor is
problematic, as functors $\J \to \C$ cannot be composed by simple
functor composition. We shall use a left Kan extension to overcome the
difficulty and obtain a skew-monoidal structure where relative monads
are skew-monoids.

\subsection{Left Kan extensions}
\label{sec:left-kan-extensions}

Left Kan extensions are one of the two canonical constructions for
extending functors. The left Kan extension along $J \in \J \to \C$
extends functors $\J \to \D$ to functors $\C \to \D$.
\[
\xymatrix@R=0.8pc@C=1pc{
& \D & \\
\J \ar@/^/[ur]^{F} \ar[rr]^{J} & & \C \ar@/_/[ul]_{\Lan_J\, F}
}
\]
It is defined as the left adjoint (if it exists) of the restriction
functor ${-} \fcomp J \in [\C,\D] \to [\J,\D]$.
%By definition, it is given by
%\begin{itemize}
%\item an object function $\Lan_J \in \obj{[\J,\D]} \to \obj{[\C,\D]}$,
%\item for any $F \in \obj{[\J,\D]}$, a map in
%  $\hom{[\J,\D]}{F}{\Lan_J\, F \cdot J}$,
%\item for any $F \in \obj{[\J,\D]}$, $G \in \obj{[\J,\D]}$, a map
%  function $\hom{[\J,\D]}{F}{G \fcomp J} \to \hom{[\C,\D]}{\Lan_J\,
%    F}{G}$
%\end{itemize}
%satisfying some coherence conditions. 
i.e., it is given by a functor $\Lan_J \in [\J,\D] \to [\C,
\D]$ and a natural isomorphism
\[
\hom{[\J,\D]}{F}{G \fcomp J} \cong \hom{[\C,\D]}{\Lan_J\, F}{G}
\]

While it is possible to work directly with this definition of left Kan
extension, we use an alternative definition, based on the coend
formula
\[
\Lan_J\, F\, X \cong \int^{Y \in \obj{\J}} \hom{\C}{J\, Y}{X} \bullet F\, Y
\]
Accordingly, we take a left Kan extension of a functor $F \in \J
\to \D$ along $J \in \J \to \C$ to be given by
\begin{itemize}
\item an object function $\Lan_J\, F \in \obj{\C} \to \obj{\D}$,
\item for any $X \in \obj{\C}$, a natural transformation \\
  $\iota_{F,X} \in
  \hom{[\J^\op,\Set]}{\hom{\C}{J\,{-}}{X}}{\hom{\D}{F\,{-}}{\Lan_J\,
      F\, X}}$,
\item for any $X \in \obj{\C}$, $Y \in \obj{\D}$ and $\theta \in
  \hom{[\J^\op,\Set]}{\hom{\C}{J\,{-}}{X}}{\hom{\D}{F\,{-}}{Y}}$, a
  map $\lcase{\theta} \in \hom{\D}{\Lan_J\, F\, X}{Y}$.
\end{itemize}
satisfying the conditions
%\begin{itemize}
%\item 
$\lcase{\theta} \circ \iota\,g =  \theta\,g$,
%\item 
$\lcase{\iota} =  \id$ 
and
%\item 
$f \circ \lcase{\theta} = \lcase{\lambda g. f \circ \theta\,g}$.
%\end{itemize}

Left Kan extensions $\Lan_J\,F\,X$ are functorial in both arguments
$F$ and $X$, i.e., $\Lan_J \in [\J,\D] \to [\C,\D]$.  For any $F \in
\obj{[\J,\D]}$, $X, Y \in \obj{\C}$, $f \in \hom{\C}{X}{Y}$,
\[
\begin{array}{l}
  \Lan_J\,F\,f \in \hom{\D}{\Lan_J\,F\,X}{\Lan_J\,F\,Y} \\
  \Lan_J\,F\,f \eqdf \lcase{\lambda g.\, \iota\,(f \comp g)}
\end{array}
\]
And for any $F, G \in \obj{[\J,\D]}$, $\tau \in \hom{[\J,\D]}{F}{G}$,
$X \in \obj{\C}$, we have
\[
\begin{array}{l}
(\Lan_J\,\tau)_X \in \hom{\D}{\Lan_J\,F\,X}{\Lan_J\,G\,X} \\
(\Lan_J\,\tau)_X \eqdf \lcase{\lambda g.\, \iota\, g \comp \tau}
\end{array}
\]
In general $\Lan_J \in [\J,\D] \to [\C,\D]$ exists, if $\J$ is small
and $\D$ is cocomplete.

\subsection{$[\J, \C] $ is skew-monoidal }
\label{sec:lax-mono}

If $\Lan_J \in [\J,\C] \to [\C,\C]$ exists, we can turn any functor
$F\in \obj{[\J,\C]}$ into one in $\obj{[\C,\C]}$. Hence we can define
a composition-like operation
\[
\begin{array}{l}
(\fcompJ) \in \obj{[\J,\C]} \times \obj{[\J,\C]} \to \obj{[\J,\C]} \\
F \fcompJ G \eqdf \Lan_J\, F\; \fcomp\; G
\end{array}
\]
This is our candidate for the tensor on $[\J,\C]$. We also need the
unital and associative laws. We define several families of maps
indexed by $X \in \obj{\C}$:
\[
\begin{array}{l}
\lambdax_X \in \hom{\C}{\Lan_J\,J\, X}{X}\\
\lambdax_X \eqdf \lcase{\lambda g.\,g} \\
(\alphay_{F,G})_X \in \hom{\C}{\Lan_J\, (F \fcomp G)\, X}{F\, (\Lan_J\,G\, X)}\\
(\alphay_{F,G})_X \eqdf \lcase{\lambda g.\,F\,(\iota\,g)} \\
(\alphax_{F,G})_X \in \hom{\C}{\Lan_J\, (\Lan_J\, F \fcomp G)\, X}{\Lan_J\, F\, (\Lan_J\,G\, X)}\\
(\alphax_{F,G})_X \eqdf (\alphay_{\Lan_J F, G})_X = \lcase{\lambda g.\, \lcase{\lambda g'.\, \iota\, (\iota\, g \comp g')}}
\end{array}
\]
All these families are natural in $X$, hence maps in $\obj{[\C,\C]}$.

From these we further define our candidate unital and associative
laws.
\[
\begin{array}{l}
\rho_F \in \hom{[\J,\C]}{F}{F \fcompJ J} \\
\rho_F \eqdf  \iota\,\id \\
\lambda_F \in \hom{[\J,\C]}{J \fcompJ F}{F} \\
\lambda_F \eqdf \lambdax \fcomp\,F \\
\alpha_{F,G,H} \in \hom{[\J,\C]}{(F\fcompJ G)\fcompJ H}{F\fcompJ(G\fcompJ H)}\\
\alpha_{F,G,H} \eqdf \alphax_{F,G} \fcomp H
\end{array}
\]

It turns out that the data so defined provide a structure that is
almost monoidal, but not quite. It is skew-monoidal in the sense of
Szlach{\'a}nyi \cite{szlachanyi}: while $\lambda$, $\rho$, $\alpha$
are generally not isomorphisms, they meet appropriate coherence
conditions, namely the conditions (a)--(e) below. Importantly, in
contrast to the properly monoidal case, all five conditions are
necessary: the conditions (a), (c), (d) do not follow from (b) and
(e).
%Monoidal-like categories where the unital
%or associative laws are not isomorphisms have been investigated before
%in the literature. 
%Our lax monoidal categories are the left
%skew-monoidal categories of Szlach{\'a}nyi \cite{szlachanyi}; a
%similar structure appeared in the work of Blute et al. \cite{blute}.

In the next section we will identify conditions on $J$ that enable us
to construct the inverses, making the skew-monoidal structure properly
monoidal.

\begin{thm}\label{thm:lax}
  If $\Lan_J \in [\J,\C] \to [\C,\C]$ exists, then
  $([\J,\C],J,\fcompJ,\lambda,\rho,\alpha)$ is a skew-monoidal
  category, i.e., $\fcompJ$ is functorial, $\lambda$, $\rho$, $\alpha$
  are natural and the following diagrams commute:
\[
\mathrm{(a)}  
\xymatrix@R=1.5pc@C=0.2pc{
    & J \fcompJ J \ar[dr]^-{\lambda_J} & \\
    J \ar[ur]^-{\rho_J} \ar@{=}[rr] & & J
    }
\quad
\mathrm{(b)}  
\xymatrix@R=1.3pc{
      (F\fcompJ J) \fcompJ G \ar[r]^{\alpha_{F,J,G}}
      & F \fcompJ (J\fcompJ G) \ar[d]^{F\fcompJ \lambda_{G}}\\
      F \fcompJ G \ar@{=}[r] \ar[u]^{\rho_{F}\fcompJ G}&  F \fcompJ G 
    }
\]
\[
\mathrm{(c)}  
\xymatrix@C=0.2pc@R=1.5pc{
  (J \fcompJ F) \fcompJ G \ar[dr]_{\lambda_F \fcompJ G} \ar[rr]^{\alpha_{J,F,G}} 
           & &  J \fcompJ (F \fcompJ G) \ar[dl]^{\lambda_{F \fcompJ G}}\\
  & F \fcompJ G & 
    }
\quad
\mathrm{(d)}  
\xymatrix@C=0.2pc@R=1.5pc{
  (F \fcompJ G) \fcompJ J \ar[rr]^{\alpha_{F,G,J}} 
           & &  F \fcompJ (G \fcompJ J) \\
  & F \fcompJ G \ar[ul]^{\rho_{F \fcompJ G}} \ar[ur]_{F \fcompJ \rho_G} & 
    }
\]
\[
\mathrm{(e)} 
\xymatrix@R=1.5pc@C=2.5pc{
(F\fcompJ (G \fcompJ H)) \fcompJ K \ar[rr]^{\alpha_{F,G \fcompJ H,K}}
  & & F\fcompJ ((G \fcompJ H)\fcompJ K) \ar[d]^{F\fcompJ \alpha_{G,H,K}}
  \\
((F\fcompJ G) \fcompJ H) \fcompJ K \ar[u]^{\alpha_{F,G,H} \fcompJ K}
      \ar[r]^{\alpha_{F\fcompJ G,H,K}}
  & (F\fcompJ G) \fcompJ (H \fcompJ K) \ar[r]^{\alpha_{F,G,H\fcompJ K}}
    & F\fcompJ (G \fcompJ (H \fcompJ K))
}
\]
\end{thm}

\proof The required properties follow from the definitions of the
functorial actions of $\Lan_J$ in both of its arguments, $\lambda$,
$\rho$, $\alpha$, and the laws of $\Lan_J$ by easy calculations.

We prove generalizations of properties (b), (c), and (e):
\[
\textrm{(b')}  
\xymatrix@R=1.3pc{
      \Lan_J (F\fcompJ J)\ar[r]^{\alphax_{F,J}}
      & F \fcompJ \Lan_J J \ar[d]^{\Lan_J F\, \lambdax}\\
      \Lan_J F \ar@{=}[r] \ar[u]^{\Lan_J \rho_{F}}&  \Lan_J F
    }
\quad
\textrm{(c')}  
\xymatrix@C=0.2pc@R=1.5pc{
  \Lan_J (J \fcompJ F)\ar[dr]_{\Lan_J \lambda_F} \ar[rr]^{\alphax_{J,F}} 
           & &  J \fcompJ \Lan_J F \ar[dl]^{\lambdax\fcomp \Lan_J F}\\
  & \Lan_J F& 
    }
\]
\quad
\[
\textrm{(e')} 
\xymatrix@R=1.5pc@C=2.5pc{
\Lan_J (F\fcompJ (G \fcompJ H))\ar[rr]^{\alphax_{F,G \fcompJ H}}
  & & F\fcompJ (\Lan_J (G \fcompJ H)) \ar[d]^{F\fcompJ \alphax_{G,H}}
  \\
\Lan_J ((F\fcompJ G) \fcompJ H)\ar[u]^{\Lan_J \alpha_{F,G,H}}
      \ar[r]^{\alphax_{F\fcompJ G,H}}
  & (F\fcompJ G) \fcompJ \Lan_J H \ar[r]^{\alpha_{F,G,\Lan_J H}}
    & F\fcompJ (G \fcompJ \Lan_J H)
}
\]

(b), (c) and (e) follow from (b'), (c') and (e') as simple instances.

We skip the proofs of functoriality of ${\fcompJ}$ and naturality of
$\lambda$, $\rho$ and $\alpha$.  The calculations for the other five
laws are as follows:

(a)
\begin{eqnarray*}
(\lambda_J)_X \comp (\rho_J)_ X 
& = & \lcase{\lambda g.\, g} \comp  {\iota\, \id}\\
& = & \id_{J\, X}
\end{eqnarray*}

(b')
\begin{eqnarray*}
\lefteqn{ \Lan_J\, F\, \lambdax_X \comp (\alphax_{F,J})_X \comp (\Lan_J\, \rho_F)_X } \\
& = & 
\Lan_J\, F\, \lambdax_X \comp (\alphax_{F,G})_X 
\comp \lcase{\lambda g.\, \iota\, g \comp \iota\, \id} \\
& = & 
\Lan_J\, F\, \lambdax_X \comp \lcase{\lambda g.\, (\alphax_{F,J})_X
     \comp \iota\, g \comp \iota\, \id} \\
& = & 
\Lan_J\, F\, \lambdax_X 
     \comp \lcase{\lambda g.\, \lcase{\lambda g.\, \lcase{\lambda g'.\, \iota\, (\iota\, g \comp g')}}
     \comp \iota\, g \comp \iota\, \id} \\
& = & 
\Lan_J\, F\, \lambdax_X 
     \comp \lcase{\lambda g.\, \lcase{\lambda g'.\, \iota\, (\iota\, g \comp g')}
     \comp \iota\, \id} \\
& = & 
\Lan_J\, F\, \lambdax_X 
     \comp \lcase{\lambda g.\, \iota\, (\iota\, g) } \\
& = & 
\lcase{\lambda g.\, \Lan_J\, F\, \lambdax_X  \comp \iota\, (\iota\, g) } \\
& = & 
\lcase{\lambda g.\, \lcase{\lambda g.\ \iota (\lcase{\lambda g'.\, g'} \comp g)}
     \comp \iota\, (\iota\, g) } \\
& = & 
\lcase{\lambda g.\, \iota\, (\lcase{\lambda g'.\, g'} \comp \iota\, g)} \\
& = & 
\lcase{\lambda g.\, \iota\, g} \\
& = & \id_{\Lan_J\, F\, X}
\end{eqnarray*}

(c')
\begin{eqnarray*}
\lambdax_{\Lan_J\, F\, X} \comp (\alphax_{J, F})_X 
& = & \lambdax_{\Lan_J\, F\, X}
\comp \lcase{\lambda g.\, \lcase{\lambda g'.\, \iota\, (\iota\, g \comp g')}} \\
& = & 
\lcase{\lambda g.\, \lambdax_{\Lan_J\, F\, X} \comp \lcase{\lambda g'.\, \iota\, (\iota\, g \comp g')}} \\
& = & 
\lcase{\lambda g.\, \lcase{\lambda g'.\, \lambdax_{\Lan_J\, F\, X} \comp \iota\, (\iota\, g \comp g')}} \\
& = & 
\lcase{\lambda g.\, \lcase{\lambda g'.\, \lcase{\lambda g''.\, g''} \comp \iota\, (\iota\, g \comp g')}} \\
& = & 
\lcase{\lambda g.\, \lcase{\lambda g'.\, \iota\, g \comp g'}} \\
& = &  
\lcase{\lambda g.\, \iota\, g \comp \lcase{\lambda g'.\ g'}} \\
& = & (\Lan_J\,  \lambda_F)_X
\end{eqnarray*}

(d)
\begin{eqnarray*}
(\alpha_{F,G,J})_X \comp (\rho_{\Lan_J\, F \fcomp G})_X 
& = & \lcase{\lambda g.\, \lcase{\lambda g'.\, \iota\, (\iota\, g \comp g')}}
  \comp \iota\, \id \\
& = & \lcase{\lambda g'.\, \iota\, (\iota\, \id \comp g')} \\
& = & \lcase{\lambda g'.\, \iota\, ((\rho_G)_X \comp g')} \\
& = & (\Lan_J\, F\, \rho_G)_X
\end{eqnarray*}

(e')
\begin{eqnarray*}
\lefteqn{ \Lan_J\, F\, (\alphax_{G,H})_X 
  \comp (\alphax_{F, \Lan_J\, G \fcomp H})_X
  \comp (\Lan_J\, \alpha_{F,G,H})_X } \\
& = & 
\Lan_J\, F\, (\alphax_{G,H})_X 
 \comp (\alphax_{F, \Lan_J\, G \cdot H})_X
 \comp \lcase{\lambda g.\, \iota\, g \comp \lcase{\lambda g'.\, \lcase{\lambda g''.\, \iota\, (\iota\, g' \comp g'')}}} \\
& = & 
\Lan_J\, F\, (\alphax_{G,H})_X 
 \comp \lcase{\lambda g.\, (\alphax_{F, \Lan_J\, G \cdot H})_X \comp \iota\, g \comp \lcase{\lambda g'.\, \lcase{\lambda g''.\, \iota\, (\iota\, g' \comp g'')}}} \\
& = & 
\Lan_J\, F\, (\alphax_{G,H})_X 
 \comp \lcase{\lambda g.\,  \lcase{\lambda g.\, \lcase{\lambda g'.\, \iota\, (\iota\, g \comp g')}} \comp \iota\, g \comp \lcase{\lambda g'.\, \lcase{\lambda g''.\, \iota\, (\iota\, g' \comp g'')}}} \\
& = & 
\Lan_J\, F\, (\alphax_{G,H})_X 
 \comp \lcase{\lambda g.\,  \lcase{\lambda g'.\, \iota\, (\iota\, g \comp g')} \comp \lcase{\lambda g'.\, \lcase{\lambda g''.\, \iota\, (\iota\, g' \comp g'')}}} \\
& = & 
\Lan_J\, F\, (\alphax_{G,H})_X 
 \comp \lcase{\lambda g.\,  \lcase{\lambda g'.\, \lcase{\lambda g''.\, \lcase{\lambda g'.\, \iota\, (\iota\, g \comp g')} \comp  \iota\, (\iota\, g' \comp g'')}}} \\
& = & 
\Lan_J\, F\, (\alphax_{G,H})_X 
 \comp \lcase{\lambda g.\,  \lcase{\lambda g'.\, \lcase{\lambda g''.\, \iota\, (\iota\, g \comp \iota\, g' \comp g'')}}} \\
& = & 
\lcase{\lambda g.\,  \lcase{\lambda g'.\, \lcase{\lambda g''.\, 
 \Lan_J\, F\, (\alphax_{G,H})_X 
 \comp  \iota\, (\iota\, g \comp \iota\, g' \comp g'')}}} \\
& = & 
\lcase{\lambda g.\,  \lcase{\lambda g'.\, \lcase{\lambda g''.\, 
 \lcase{\lambda g.\, \iota\, ((\alphax_{G,H})_X \comp g)}
 \comp  \iota\, (\iota\, g \comp \iota\, g' \comp g'')}}} \\
& = & 
\lcase{\lambda g.\,  \lcase{\lambda g'.\, \lcase{\lambda g''.\, 
 \iota\, ((\alphax_{G,H})_X \comp \iota\, g \comp \iota\, g' \comp g'')}}}\\
& = & 
\lcase{\lambda g.\,  \lcase{\lambda g'.\, \lcase{\lambda g''.\, 
 \iota\, (\lcase{\lambda g.\, \lcase{\lambda g'.\, \iota\, (\iota\, g \comp g')}} \comp \iota\, g \comp \iota\, g' \comp g'')}}}\\
& = & 
\lcase{\lambda g.\,  \lcase{\lambda g'.\, \lcase{\lambda g''.\, 
 \iota\, (\lcase{\lambda g'.\, \iota\, (\iota\, g \comp g')} \comp \iota\, g' \comp g'')}}}\\
& = &
\lcase{\lambda g.\, \lcase{\lambda g'.\, \lcase{\lambda g''.\, \iota\, (\iota\, (\iota\, g \comp g') \comp g'')}}} \\
& = & 
\lcase{\lambda g.\, \lcase{\lambda g'.\, \lcase{\lambda g.\, \lcase{\lambda g''.\, \iota\, (\iota\, g \comp g'')}} \comp \iota\, (\iota\, g \comp g')}} \\
& = & 
\lcase{\lambda g.\, \lcase{\lambda g'.\, (\alphax_{F, G})_{\Lan_J\, H\, X}  \comp \iota\, (\iota\, g \comp g')}} \\& = & 
(\alphax_{F, G})_{\Lan_J\, H\, X} 
 \comp \lcase{\lambda g.\, \lcase{\lambda g'.\, \iota\, (\iota\, g \comp g')}} \\
& = & 
(\alphax_{F, G})_{\Lan_J\, H\, X} 
  \comp (\alphax_{\Lan_J\, F \cdot G, H})_X 
\end{eqnarray*}

\begin{exa}
  The functor category $[\J, \C]$ is skew-monoidal, but not monoidal,
  for $\J \eqdf \Set$, $\C \eqdf \Set$, $J\, X \eqdf X \times S$.

  In this case, we have $\Lan_J\, F\, X \cong \int^Y ((Y \times S \to
  X) \times F\, Y) \cong \int^Y ((Y \to X^S) \times F\, Y) \cong F\,
  (X^S)$.

  Accordingly, $(\rho_F)_X \in F\, X \to \Lan_J\, F\, (J\, X)$ is
  given by $(\rho_X)_F \eqdf F\, \mathrm{coeval}_X \in F\, X \to F((X
  \times S)^S)$, $\lambdax_X \in \Lan_J\, J\, X \to X$ is given by
  $\lambdax_X \eqdf \mathrm{eval}_X \in X^S \times S \to X$. It is clear
  that $\rho$ and $\lambdax$ are not isomorphisms in this case.

  The map $(\alphay_{F, G})_X \in \Lan_J\, (F \fcomp G)\, X \to
F\, (\Lan_J\, G\, X)$ however is given by the identity on $F\,(G\,
(X^S))$ and is therefore trivially an isomorphism.
\end{exa}

\begin{exa}
  For $\J \eqdf \Set$, $\C \eqdf \Set$, $J\, X \eqdf X + E$, the
  functor category $[\J, \C]$ is also skew-monoidal. But in this case,
  even the associativity law $\alpha$ fails to be an isomorphism.

  We have $\Lan_J\, F\, X \cong \int^Y ((Y +E \to X) \times F\, Y)
  \cong \int^Y ((Y \to X) \times (E \to X) \times F\, Y) \cong F\, X
  \times X^E$. 

  Accordingly, $\rho$, $\lambda$, $\alpha$ are the canonical natural
  transformations with components $(\rho_F)_X \in F\, X \to F(X +E)
  \times (X+E)^E$, $\lambdax_X \in (X + E) \times X^E \to X$,
  $(\alphay_{F,G})_X \in F\, (G\, X) \times X^E \to F\, (G\, X \times
  X^E)$.  None of these has an inverse.
\end{exa}
\newpage

\subsection{Relative monads are the same as skew-monoids in $[\J, \C]$}
\label{sec:constr-monoid}

With a skew-monoidal structure present on the functor category $[\J,\C]$, 
we should expect that relative monads on $J$ are the same thing
as skew-monoids in this structure, generalizing the case of ordinary
monads on $\C$ and the strict monoidal structure on the endofunctor
category $[\C, \C]$. This is indeed the case.

\begin{thm}\label{thm:monoid}
Assume that $\Lan_J \in [\J,\C] \to [\C,\C]$ exists. 
\begin{enumerate}
\item Given a relative monad $(T, \eta, (-)^*)$ on $J$, define, for
  any $X \in \obj{\J}$, a map $\mu_X \in
  \hom{\C}{\Lan_J\,T\,(T\,X)}{T\,X}$ by $\mu_X \eqdf \lcase{(-)^*}$. This is well-defined, since $(-)^*$ is
  natural: $(-)^* \in \hom{[\J^\op,
    \Set]}{\hom{\C}{J\,{-}}{T\,X}}{\hom{\C}{T\,{-}}{T\,X}}$.

  Then $(T, \eta, \mu)$ is a skew-monoid in the skew-monoidal
  category $([\J,\C], J, \fcompJ, \lambda, \rho, \alpha)$: we have
  that $T \in \obj{[\J,\C]}$, $\eta \in \hom{[\J,\C]}{J}{T}$ and $\mu
  \in \hom{[\J,\C]}{T \fcompJ T}{T}$, and the following diagrams
  commute in $[\J,\C]$:
\[
\xymatrix@C=1pc@R=1.5pc{
T \fcompJ J \ar[rr]^{T \fcompJ \eta} & & T \fcompJ T \ar[dd]^{\mu} \\
T \ar@{=}[drr] \ar[u]^-{\rho_T} & & \\
& & T
}
\quad
\xymatrix@C=1pc@R=4pc{
J \fcompJ T \ar[r]^-{\lambda_T} \ar[d]^{\eta \fcompJ T}
  & T \ar@{=}[dr] & \\
T \fcompJ T \ar[rr]^{\mu} & & T
}
\quad
\xymatrix@C=0.5pc@R=1.2pc{
& T \fcompJ (T \fcompJ T) \ar[rr]^-{T \fcompJ \mu} & & T \fcompJ T \ar[dd]^{\mu} \\
(T \fcompJ T) \fcompJ T \ar[ur]^-{\alpha_{T,T,T}} \ar[d]_{\mu \fcompJ T} & & \\
T \fcompJ T \ar[rrr]^{\mu} & & & T 
}
\]

% \[
% \xymatrix{J\fcompJ T\ar[r]^{\eta\fcompJ T}\ar[rd]_{\lambda_T}& 
%       T\fcompJ T \ar[d]_{\mu}&
%       T \fcompJ J\ar[l]^{T\fcompJ\eta}\\
%       &T \ar@{=}[r] & T \ar[u]^{\rho_T}}
% \;\xymatrix{(T\fcompJ T)\fcompJ T
%       \ar[r]^{\alpha_{T, T, T}}
%       \ar[d]_{\mu\fcompJ T}&
%       T\fcompJ (T\fcompJ T)
%       \ar[r]^{T\fcompJ\mu}&
%       T\fcompJ T\ar[d]^\mu\\
%       T\fcompJ T\ar[rr]^\mu&&
%       T
%     }
% \]
\item Given a skew-monoid $(T, \eta, \mu)$ in $([\J,\C], J, \fcompJ,
  \lambda, \rho, \alpha)$, define, for any $X, Y \in \obj{\J}$, a
  function $(-)^* \in \hom{\C}{J\, X}{T\, Y} \to \hom{\C}{T\, X}{T\,
    Y}$ by $k^* \eqdf \mu_Y \comp \iota\, k$.  Then $(T, \eta, (-)^*)$
  is a relative monad on $J$.

\item The above correspondence is bijective.
\end{enumerate}
\end{thm}

\proof\hfill
\begin{enumerate}
\item The required properties follow from the definitions of
$\mu$ and the functorial action of $\Lan_J$ and from $T$ being a
relative monad by the laws of $\Lan_J$ alone.

For naturality of $\mu$, we easily verify that, for any $f \in
\hom{J}{X}{Y}$,
\begin{eqnarray*}
T\, f \comp \mu_X 
& = & T\, f \comp \lcase{\lambda g.\, g^*} \\
& = & \lcase{\lambda g.\, T\, f \comp g^*} \\
& = & \qquad \mbox{\{~by naturality of $(-)^*$~\}} \\
&   & \lcase{\lambda g.\, (T\, f \comp g)^*} \\
& = & \lcase{\lambda g.\, \lcase{\lambda g.\, g^*} \comp \iota\,(T\, f \comp g)} \\
& = & \lcase{\lambda g.\, g^*} \comp \lcase{\lambda g.\, \iota\, (T\, f \comp g)} \\
& = & \mu_Y \comp \Lan_J\, T\, (T\, f) 
\end{eqnarray*}

The right unital law of $T$ as a monoid is verified by
\begin{eqnarray*} 
\mu_X \comp (\Lan_J\, \eta)_{T\, X} 
& = & \lcase{\lambda g.\, g^*} \comp \lcase{\lambda g.\ \iota\, g \comp \eta} \\
& = & \lcase{\lambda g.\ \lcase{\lambda g.\, g^*} \comp \iota\, g \comp \eta} \\
& = & \lcase{\lambda g.\, g^* \comp \eta} \\
& = & \qquad \mbox{\{~by right unital law of $T$ as a relative monad~\}} \\
&   & \lcase{\lambda g.\, g} \\
& = & \lambda_{T,X} 
\end{eqnarray*}

The left unital law of $T$ as a monoid is checked by
\begin{eqnarray*}
\mu_X \comp \Lan_J\, T\, \eta_X \comp (\rho_{T})_X 
& = & \lcase{\lambda g.\, g^*} \comp \lcase{\lambda g.\, \iota\, (\eta_X \comp g)} \comp (\rho_{T})_X \\
& = & \lcase{\lambda g.\, \lcase{\lambda g.\, g^*} \comp \iota\, (\eta_X \comp g)} \comp (\rho_{T})_X \\
& = & \lcase{\lambda g.\, (\eta_X \comp g)^*} \comp (\rho_{T})_X \\
& = & \lcase{\lambda g.\, (\eta_X \comp g)^*} \comp \iota\, \id_{J\, X} \\
& = & (\eta_X \comp \id_{J\, X})^* \\
& = & \qquad \mbox{\{~by left unital law of $T$ as a relative monad~\}} \\
&   & \id_{T\, X} 
\end{eqnarray*}

The associativity of $T$ as a monoid is verified by
\begin{eqnarray*}
 \mu_X \comp \Lan_J\, T\, \mu_X \comp (\alpha_{T,T,T})_X 
& = & \lcase{\lambda g.\, g^*} \comp \lcase{\lambda g.\, \iota\, (\mu_X \comp g)} \comp (\alpha_{T,T,T})_X \\
& = & \lcase{\lambda g.\, \lcase{\lambda g.\, g^*} \comp \iota\, (\mu_X \comp g)} \comp (\alpha_{T,T,T})_X \\
& = & \lcase{\lambda g.\, (\mu_X \comp g)^*} \comp (\alpha_{T,T,T})_X \\
& = & \lcase{\lambda g.\, (\mu_X \comp g)^*} \comp \lcase{\lambda g.\, \lcase{\lambda g'.\, \iota\, (\iota\, g \comp g')}} \\
& = & \lcase{\lambda g.\, \lcase{\lambda g.\, (\mu_X \comp g)^*} \comp \lcase{\lambda g'.\, \iota\, (\iota\, g \comp g')}} \\
& = & \lcase{\lambda g.\, \lcase{\lambda g'.\, \lcase{\lambda g.\, (\mu_X \comp g)^*} \comp \iota\, (\iota\, g \comp g')}} \\
& = &  \lcase{\lambda g.\, \lcase{\lambda g'.\, (\mu_X \comp \iota\, g \comp g')^*}}\\
& = & \lcase{\lambda g.\, \lcase{\lambda g'.\, (\lcase{\lambda g.\, g^*} \comp \iota\, g \comp g')^*}}\\
& = & \lcase{\lambda g.\, \lcase{\lambda g'.\, (g^* \comp g')^*}}\\
& = &      \quad \mbox{\{~by associative law of $T$ as a relative monad~\}} \\
&   & \lcase{\lambda g.\, \lcase{\lambda g'.\, g^* \comp g'^*}} \\
& = & \lcase{\lambda g.\, g^* \comp \lcase{\lambda g'.\, g'^*}} \\
& = & \lcase{\lambda g.\, g^* \comp \mu} \\
& = & \lcase{\lambda g.\, \lcase{\lambda g.\, g^*} \comp \iota\, g \comp \mu} \\
& = & \lcase{\lambda g.\, g^*} \comp \lcase{\lambda g.\, \iota\, g \comp \mu} \\
& = & \mu_X \comp (\Lan_J\, \mu)_{T\, X}
\end{eqnarray*}

\item The claim follows from the definitions of $(-)^*$ and the
functorial action of $\Lan_J$ and from $T$ being a skew-monoid by the
laws of $\Lan_J$.

\item The claim follows from the definitions of $\mu$ and $(-)^*$ from
each other and the laws of $\Lan_J$.
\qed
\end{enumerate}

\noindent The bijective correspondence between relative monads on $J$ and
skew-monoids in $[\J, \C]$ extends to an isomorphism of categories.

\begin{thm}\label{thm:monoid-mor}
Assume that $\Lan_J \in [\J,\C] \to [\C,\C]$ exists. 
\begin{enumerate}
\item A morphism $\sigma$ between relative monads $(T, \eta, (-)^*)$
  and $(T', \eta', (-)^{*'})$ is a morphism between the corresponding
  skew-monoids $(T, \eta, \mu)$ and $(T, \eta', \mu')$: we have that
  $\sigma \in [\J,\C](T, T')$ and the following diagrams commute in
  $[\J,\C]$:
\[
\xymatrix@R=0.9pc@C=3pc{
 & T \ar[dd]^{\sigma} \\
J \ar[ur]^{\eta} \ar[dr]_{\eta'} 
 & \\
 & T'
}
\qquad
\xymatrix@R=2.8pc@C=3pc{
T \fcompJ T \ar[r]^{\mu} \ar[d]_{\sigma \fcompJ \sigma} 
 & T \ar[d]^{\sigma} \\
T' \fcompJ T' \ar[r]^{\mu'} 
 & T'
}
\]

\item A morphism $\sigma$ between skew-monoids $(T, \eta, \mu)$ and
  $(T', \eta', \mu')$ is also a morphism between the corresponding
  relative monads $(T, \eta, (-)^*)$ and $(T, \eta', (-)^{*'})$.

\item The above correspondence is an isomorphism of the categories of
  relative monads on $J$ and skew-monoids in the skew-monoidal category
  $([\J,\C], J, \fcompJ, \lambda, \rho, \alpha)$.
  
\end{enumerate}
\end{thm}

\proof\hfill 
\begin{enumerate}
\item We have already observed that $\sigma$ is natural. The unit
  preservation law for $\sigma$ as a skew-monoid morphism is the same
  as the unit preservation law of $\sigma$ as a relative monad
  morphism.

  The multiplication preservation law of $\sigma$ as a skew-monoid
  morphism follows from the definitions of $\mu$, $\mu'$ from $(-)^*$,
  $(-)^{*'}$ and Kleisli extension preservation of $\sigma$ as a
  relative monad morphism by the laws of $\Lan_J$:
\begin{eqnarray*}
\lefteqn{ \mu'_X \comp (\Lan_J\, \sigma)_{T'\, X} \comp \Lan_J\, T\, \sigma_X } \\
& = &  
\mu'_X \comp (\Lan_J\, \sigma)_{T'\, X} \comp \lcase{\lambda g.\, \iota\, (\sigma_X\, \comp g)} \\
& = & 
\mu'_X \comp \lcase{\lambda g.\, (\Lan_J\, \sigma)_{T'\, X} \comp \iota\, (\sigma_X\, \comp g)} \\
& = & 
\mu'_X \comp \lcase{\lambda g.\, \lcase{\lambda g.\, \iota\, g \comp \sigma} \comp \iota\, (\sigma_X\, \comp g)} \\
& = & 
\mu'_X \comp \lcase{\lambda g.\, \iota\, (\sigma_X \comp g) \comp \sigma} \\
& = & 
\lcase{\lambda g.\, \mu'_X \comp \iota\, (\sigma_X \comp g) \comp \sigma} \\
& = & 
\lcase{\lambda g.\, \lcase{\lambda g.\, g^{*'}} \comp \iota\, (\sigma_X \comp g) \comp \sigma} \\
& = &
\lcase{\lambda g.\, (\sigma_X \comp g)^{*'} \comp \sigma} \\
& = & \qquad \mbox{\{~by Kl. ext. pres. law of $\sigma$ as rel. mon. morphism~\}} \\
&   & \lcase{\lambda g.\, \sigma_X \comp g^*} \\
& = & 
\sigma_X \comp \lcase{\lambda g.\, g^*} \\
& = & 
\sigma_X \comp \mu_X
\end{eqnarray*}

\item The claim follows from the definitions of $(-)^*$, $(-)^{*'}$
  from $\mu$, $\mu'$, the unit and multiplication preservation of
  $\sigma$ as a skew-monoid morphism and the laws of $\Lan_J$.
\item The claim follows from the mutual definitions of $\mu$, $\mu'$
  from $(-)^*$, $(-)^{*'}$ by the laws of $\Lan_J$.
\qed
\end{enumerate}

\noindent We have seen that, in the presence of $\Lan_J$, relative monads can be
defined equivalently in the Kleisli extension and multiplication based
formats. Restriction $(-)^\flat$ is defined for multiplication as
follows. Given a monad $(T, \eta, \mu)$, with $\mu \in [\C, \C](T
\fcomp T, T)$, we set
$\mu^\flat \eqdf 
\xymatrix{
\Lan_J\, (T \fcomp J) \fcomp T \fcomp J 
     \ar[r]^-{\alphay_{T,J} \fcomp T \fcomp J}
  & T \fcomp \Lan_J\, J \fcomp T \fcomp J
     \ar[r]^-{T \fcomp \lambdax \fcomp T \fcomp J}
    & T \fcomp T \fcomp J
     \ar[r]^-{\mu \fcomp J}
      & T \fcomp J
}
$

\subsection{An equivalent version of EM-algebras}
\label{sec:alt-EM}

Just as the availability of $\Lan_J \in [\J,\C] \to [\C,\C]$
allows us to define relative monads based on $\mu$ rather than
$(-)^*$, it also facilitates a more traditional-style definition of
EM-algebras.

\begin{defi}
 If $\Lan_J \in [\J,\C] \to [\C,\C]$ exists, an EM-algebra$\alt$ of a
 relative monad $T$ on $J$ is given by an object $X \in \obj{\C}$ and
 a map $x \in \hom{\C}{\Lan_J\, T\, X}{X}$, making the following
 diagrams commute in $\C$:
\[
\xymatrix@C=1pc@R=4.2pc{
\Lan_J\, J\, X\ar[r]^-{\lambdax_X} \ar[d]^{(\Lan_J\, \eta)_X}
 & X \ar@{=}[dr] & \\
\Lan_J\, T\, X \ar[rr]^{x} & & X
}
\quad
\xymatrix@C=0.7pc@R=1.4pc{
& \Lan_J\, T\, (\Lan_J\, T\, X) \ar[rr]^-{\Lan_J\, T\, x} & & \Lan_J\, T\, X \ar[dd]^{x} \\
\Lan_J\, (\Lan_J\, T \fcomp T)\, X  \ar[ur]^-{(\alphax_{T,T})_X} \ar[d]_{(\Lan_J\, \mu)_X} & & \\
\Lan_J\, T\, X \ar[rrr]^{x} & & & X 
}
\]   
 An EM-algebras$\alt$ map between $(X,x)$, $(Y,y)$ is a map $h \in
 \hom{\C}{X}{Y}$, making the following diagram commute in $\C$:
\[
\xymatrix@R=2.8pc@C=3pc{
\Lan_J\, T\, X \ar[r]^{\Lan_J\, T\, h} \ar[d]_{x} 
 & \Lan_J\, T\, Y \ar[d]^{x} \\
X \ar[r]^{h} 
 & Y
}
\]
\end{defi}

EM-algebra $\alt$ and EM-algebra$\alt$ maps of $T$ form a category
$\EMalt{T}$ that inherits its identities and composition from $\C$.

\begin{thm}
 Assume that $\Lan_J \in [\J,\C] \to [\C,\C]$ exists. Consider a
 relative monad $T$ on $J$.
\begin{enumerate}
\item An EM-algebra $(X, \chi)$ gives rise to an EM-algebra$\alt$ $(X,
 \lcase{\chi})$. 
\item An EM-algebra$\alt$ $(X, x)$ gives rise to an EM-algebra $(X,
 \lambda g.\ x \comp \iota\, g)$. 
\item This correspondence is a bijection.
\item An EM-algebra map $h$ between $(X, \chi)$, $(Y, \psi)$
 is also an EM-algebra$\alt$ map between $(X, \lcase{\chi})$, $(Y,
 \lcase{\psi})$.
\item An EM-algebra$\alt$ map $h$ between
 $(X, x)$, $(Y, y)$ is also an EM-algebra map between $(X,
 \lambda g.\ x \comp \iota\, g)$, $(Y, \lambda g.\ y \comp \iota\,
 g)$.
\item The categories $\EM{T}$ and $\EMalt{T}$ are isomorphic.
\end{enumerate}
\end{thm}

\proof
We only prove (1) and (4).

\begin{enumerate}
\item The two EM$\alt$-algebra laws of $(X, \lcase{\chi})$ are
  obtained from the definitions of $\lambdax$, $\mu$ and the laws of
  $\Lan_J$ with the help of the corresponding EM-algebra laws of $(X,
  \chi)$ as follows:
\begin{eqnarray*}
\lcase{\chi} \comp (\Lan_J\, \eta)_X 
& = & 
\lcase{\chi} \comp \lcase{\lambda g.\, \iota\, g \comp \eta} \\
& = & 
\lcase{\lambda g.\, \lcase{\chi} \comp \iota\, g \comp \eta} \\
& = & 
\lcase{\lambda g.\, \chi\, g \comp \eta} \\
& = & 
\qquad \mbox{\{~by 1st EM-algebra law~\}} \\
&   & 
\lcase{\lambda g.\, g} \\
& = & 
\lambdax_X 
\end{eqnarray*}

\begin{eqnarray*}
\lcase{\chi} \comp (\Lan_J\, \mu)_X 
& = & 
\lcase{\chi} \comp \lcase{\lambda g.\, \iota\, g \comp \mu} \\
& = & 
\lcase{\lambda g.\, \lcase{\chi} \comp \iota\, g \comp \mu} \\
& = & 
\lcase{\lambda g.\, \chi\, g \comp \mu} \\
& = & 
\lcase{\lambda g.\, \chi\, g \comp \lcase{\lambda g'.\, g'^*}} \\
& = &
\lcase{\lambda g.\, \lcase{\lambda g'.\, \chi\, g \comp g'^*}} \\
& = & \qquad \mbox{\{~by 2nd EM-algebra law~\}} \\ 
&   & 
\lcase{\lambda g.\, \lcase{\lambda g'.\, \chi\, (\chi\, g \comp g')}} \\
& = & 
\lcase{\lambda g.\, \lcase{\lambda g'.\, \lcase{\chi} \comp \iota\, (\chi\, g \comp g')}} \\
& = & 
\lcase{\lambda g.\, \lcase{\chi} \comp \lcase{\lambda g'.\, \iota\, (\chi\, g \comp g')}} \\
& = & 
\lcase{\chi} \comp \lcase{\lambda g.\, \lcase{\lambda g'.\, \iota\, (\chi\, g \comp g')}} \\
& = & 
\lcase{\chi} \comp \lcase{\lambda g.\, \lcase{\lambda g'.\, \iota\, (\lcase{\chi} \comp \iota\, g \comp g')}} \\
& = & 
%\lcase{\chi} \comp \lcase{\lambda g.\, \lcase{\lambda g'.\, \iota\, (\lcase{\chi} \comp \iota\, g \comp g')}} \\
%& = & 
\lcase{\chi} 
\comp \lcase{\lambda g.\, \lcase{\lambda g'.\, \lcase{\lambda g.\, \iota\, (\lcase{\chi} \comp g)} \comp \iota\, (\iota\, g \comp g')}} \\ 
& = & 
\lcase{\chi} 
\comp \lcase{\lambda g.\, \lcase{\lambda g'.\, \Lan_J\, T\, \lcase{\chi} \comp \iota\, (\iota\, g \comp g')}} \\
%& = & 
%\lcase{\chi} 
%\comp \lcase{\lambda g.\, \lcase{\lambda g'.\, \Lan_J\, T\, \lcase{\chi} \comp \iota\, (\iota\, g \comp g')}} \\
& = & 
\lcase{\chi} 
\comp \lcase{\lambda g.\, \Lan_J\, T\, \lcase{\chi} \comp \lcase{\lambda g'.\, \iota\, (\iota\, g \comp g')}} \\
& = & 
\lcase{\chi} \comp \Lan_J\, T\, \lcase{\chi} 
\comp \lcase{\lambda g.\, \lcase{\lambda g'.\, \iota\, (\iota\, g \comp g')}} \\
& = & 
\lcase{\chi} \comp \Lan_J\, T\, \lcase{\chi} \comp (\alphax_{T,T})_X
\end{eqnarray*}

\setcounter{enumi}{3}
\item The EM$\alt$-algebra map law of $h$ is obtained from the laws of
  $\Lan_J$ with the help of the EM-algebra map law of of $h$ as
  follows:
\begin{eqnarray*}
h \comp \lcase{\chi} 
& = & \lcase{\lambda g.\, h \comp \chi\, g} \\
& = & \qquad \mbox{\{~by EM-algebra morphism law~\}} \\
&   & \lcase{\lambda g.\, \upsilon\, (h \comp g)} \\
& = & \lcase{\lambda g.\, \lcase{\upsilon} \comp \iota\, (h \comp g)} \\
& = & \lcase{\upsilon} \comp \lcase{\lambda g.\, \iota\, (h \comp g)} \\
& = & \lcase{\upsilon} \comp \Lan_J\, T\, h
\end{eqnarray*}
\vspace{-30pt}

\qed
\end{enumerate}

% ======================================================================
% ======================================================================
% ======================================================================

\section{Well-behaved relative monads}
\label{sec:proper-monoids}

It is somewhat unsatisfactory to obtain that $[\J, \C]$ is just
skew-monoidal, rather than properly monoidal. This begs the question:
would some conditions on $J$ ensure a properly monoidal structure? The
answer is yes. Mild conditions turn the skew-monoidal structure of
$[\J, \C]$ into properly monoidal. What is more, the same conditions
also allow relative monads on $J$ to extend to monads on $\C$.

\subsection{Well-behavedness conditions}

We define three well-behavedness conditions on $J$. They are
additional to the existence of $\Lan_J \in [\J,\C] \to [\C,\C]$ and
require the constituent maps of three canonical families, which are
actually natural, to be isomorphisms. 

\begin{defi}
  $J \in \J \to \C$ is \emph{well-behaved}, if not only does $\Lan_J
  \in [\J,\C] \to [\C,\C]$ exist, but also the following three
  conditions hold:

\begin{enumerate}

\item $J$ is \emph{fully faithful}, i.e., for any $X, Y \in \obj{\J}$,
  there is an inverse to the map
\[
\begin{array}{l}
J_{X,Y} \in \hom{\J}{X}{Y} \to \hom{\C}{J\, X}{J\, Y} \\
J_{X,Y}\, f \eqdf J\, f
\end{array}
\]

\item $J$ is \emph{dense}, i.e., for any $X, Y \in \obj{\C}$, there is
  an inverse to the map
\[
\begin{array}{l}
    K_{X,Y} \in \hom{\C}{X}{Y} \to \hom{[\J^\op,\Set]}{\hom{\C}{J\,{-}}{X}}{\hom{\C}{J\,{-}}{Y}}\\
    K_{X,Y}\,f \eqdf \lambda g .\, f \circ g
\end{array}
\]
This is the same as to say that the associated \emph{nerve} functor $K \in \C \to [\J^\op, \Set]$, defined by $K\, X \eqdf \hom{\C}{J\, {-}}{X}$, is fully faithful.

%\end{enumerate}

%\begin{enumerate}
%\setcounter{enumi}{2}

\item For any $F \in \J \to \C$, $X \in \obj{\J}$, $Y \in \obj{\C}$,
  there is an inverse to the map
\[
\begin{array}{l}
    L^F_{X,Y} \in \Lan_J\, (\hom{\C}{J\, X}{F {-}})\, Y \to \hom{\C}{J\,X}{\Lan_J\,F\,Y}  \\
    L^F_{X,Y} \eqdf \lcase{\lambda g.\, \lambda g'.\, \iota\, g \comp g'}
\end{array}
\]
This condition says that the nerve functor $K$ preserves left Kan
extensions of functors $F \in \J \to \C$ along $J$.
\end{enumerate}
%(Notice that the domain of $L^F_{X,Y}$ is given by the left Kan
%extension of a functor $\J \to \Set$.)
\end{defi}

% % For our purposes, these conditions are mild. 
% \begin{exa}
%   The functor $\Jfin \in \Fin \to \Set$ is well-behaved. The functor
%   $\JU \in \UU \to \Cat$ of Example~\ref{ex:ic} is well-behaved, if
%   the type-theoretic universe $\U\in\Set,\El\in \U\to\Set$ is is closed
%   under dependent products (categorically this corresponds to the
%   induced category $\UU$ being cartesian).
% \end{exa}

\noindent The functors $\Jfin \in \Fin \to \Set$ and $\JU \in \UU \to \Set$ we
have considered in our examples \ref{ex:qc}, \ref{ex:ulam}
resp.\ \ref{ex:ic} turn out to be well-behaved. This is a consequence
of a general construction. 

Let $\U \in \obj{\Set} $ and $\El \in \U \to \obj{\Set}$ be a
type-theoretic universe (as in Example~\ref{ex:ic}).  As we already
explained above, $\U$ and $\El$ define a category $\UU$ by $\obj{\UU}
\eqdf \U$ and $\hom{\UU}{A}{B} \eqdf \El\, A \to \El\, B$ and a
functor $\JU \in \UU \to \Set$ by $\JU\, A \eqdf \El\, A$ on objects
and $\JU\, f \eqdf f$ on maps. In order for $\JU$ to be well-behaved,
it suffices that the universe has dependent products, i.e., that we
have
\begin{align*} 
& \one \in \U \\ 
& \sigma \in \Pi A \in \U .\, (\El\, A \to \U)\to \U
\end{align*}
such that 
\begin{eqnarray*}
  \El\,\one & = & 1 \\
  \El\,(\sigma\,A\,B) & = & \Sigma a \in \El\,A .\, \El\,(B\,a)
\end{eqnarray*}

%We are exploiting the fact that ends and coends in $\Set$ can be
%concretely constructed. Given a small category $\J$ and a bifunctor
%$F \in \J^\op \to[\J,\Set]$ we define:
%\begin{description}
%\item[ends] 
%\[ \int_{X\in\obj{\J}}\,F\,X\,X = \{ \alpha \in \Pi X \in \obj{\J} .\, F\,X\,X %\mid \phi\,\alpha \}\]
%where $\phi\,\alpha$ expresses the condition that for any $X,Y\in\obj{\J}, f\in \hom{\J}{X}{Y}$,
%the equation $F\,X\,f\,(\alpha\,X) = F\,f\,Y\,(\alpha\,Y)$ holds in $F\,X\,Y$.
%\item[coends] 
%\[ \int^{X\in\obj{\J}}\,F\,X\,X = \Sigma X \in \obj{\J} .\, F\,X\,X / \sim \]
%where $\sim$ is the least equivalence relation which for any $X,Y\in\obj{\J}, f\in \hom{\J}{X}{Y}$ and $z\in F\,X\,Y$ contains
%\[ (X, F\,X\,f\,z) \sim (Y,F\,f\,Y\,z) \] 
%\end{description}

\noindent To prove this, we exploit the fact that (small) coends in $\Set$ can
be constructed explicitly. Given any small category $\J$ and functors
$J, F \in \J \to \Set$, for any $X \in \obj{\Set}$, we have
\begin{eqnarray*}
\Lan_J\,F\,X & = & \int^{Z \in \obj{\J}} \hom{\Set}{J\, Z}{X} \bullet F\, Z \\
& \cong & (\Sigma Z \in \obj{\J}.\, (J\, Z \to X) \times F\,Z) / {\sim_X}
\end{eqnarray*}
where $\sim_X$ is the least equivalence relation containing $(Z,g \circ
J\,h, x) \sim_X (W,g,F\,h\,x)$ for any $g \in J\,W \to X, x \in F\,Z,
h\in \hom{\J}{Z}{W}$. We can derive a specialized definition of $L$ in
this case:
\begin{align*}
  & L^F_{X,Y} \in \Lan_J\, (J\,X \to F\,-)\,Y \to (J\,X \to \Lan_J\,F\,Y) \\
  & L^F_{X,Y}\,(Z,g, k) \eqdf \lambda x.\, (Z , g , k\, x)
\end{align*}
We omit the verification that the equivalence is preserved in the
definition of $L^F_{X,Y}$. 

We can now show that inverses to $J$, $K$, $L$ exist and hence $\JU$
is well-behaved.

\begin{thm}
For any type-theoretic universe closed under dependent products,
the functor $J \eqdf \JU \in \UU \to\Set$ is well-behaved:
\begin{enumerate}
\item For any $A,B\in \obj{\UU}$, $J_{A,B}$ has an inverse
  \[{J_{A,B}}^{-1} \in (\El\,A \to \El\,B) \to (\El\, A \to \El\, B)\]

\item  For any $X,Y\in \obj{\Set}$, $K_{X,Y}$ has an inverse
\[{K_{X,Y}}^{-1} \in (\int_{C \in \obj{\UU}} (\El\,C \to X) \to (\El\,C \to Y)) \to (X \to Y)\]

\item For any $F \in \UU \to\Set$, $A \in \obj{\UU}$, $Y \in
  \obj{\Set}$, $L^F_{A,Y}$ has an inverse
  \[{L^F_{A,Y}}^{-1} \in (\El\,A \to \Lan_J\,F\,Y) \to \Lan_J\, (\El\,A\to F\,-)\,Y\]

\end{enumerate} 
\end{thm}

\proof
\hfill
\begin{enumerate}

\item This is obvious, since $J_{A,B}\, f = J\, f = f$.

\item Given $\tau \in \int_{C \in\obj{\UU}} (\El\, C \to X) \to (\El\,
  C \to Y)$, we can construct $K_{X,Y}^{-1}\, \tau \in X \to Y$ as
\[ 
{K_{X,Y}}^{-1}\,\tau \eqdf \lambda x.\, \tau\, (\lambda z.\, x)\, \ast 
\] 
  where $\ast$ is the unique element of $\El\, \one = 1$. We
  verify that this is indeed the inverse:

 \begin{eqnarray*}
 {K_{X,Y}}^{-1}\,(K_{X,Y}\,f)
    %& = & \lambda x.\, (K_{X,Y}\,f)\, (\lambda z.\, x)\, \ast \\
    & = & {K_{X,Y}}^{-1}\, (\lambda g.\, f \comp g) \\
    & = & \lambda x.\, (f \comp (\lambda z.\, x))\, \ast \\
    & = & \lambda x.\, f\, ((\lambda z.\, x)\, \ast) \\
    & = & \lambda x.\, f\,x  \\
    & = & f
  \end{eqnarray*}
  
  \begin{eqnarray*}
  K_{X,Y}\,({K^{-1}}_{X,Y}\,\tau)  
    %& = & \lambda g.\,  {K_{X,Y}}^{-1}\,\tau \comp g \\
    & = &  K_{X,Y}\,(\lambda x'.\, \tau\, (\lambda z.\, x')\, \ast)  \\
    & = & \lambda g.\, (\lambda x'.\, \tau\, (\lambda z.\, x')\, \ast) \comp g \\
%    & = & \lambda g.\, \lambda x.\, {K_{X,Y}}^{-1}\,\tau\,(g\,x) \\
    & = & \lambda g.\, \lambda x.\, (\lambda x'.\, \tau\, (\lambda z.\, x')\,\ast)\,(g\, x) \\
    & = & \lambda g.\, \lambda x.\, \tau\, (\lambda z.\, g\,x)\, \ast\\
    & = & \lambda g.\, \lambda x.\, \tau\, (g \comp (\lambda z.\, x))\, \ast\\ 
    & = & \qquad \{\textrm{~by naturality of $\tau$~}\} \\
    &   & \lambda g.\, \lambda x.\, (\tau\, g \comp (\lambda z.\, x))\, \ast\\
    & = & \lambda g.\, \lambda x.\, \tau\, g\, ((\lambda z.\, x)\, \ast)\\
    & = & \lambda g.\, \lambda x.\, \tau\, g\, x \\
    & = & \tau
  \end{eqnarray*}

\item Given $f\in \El\, A \to \Lan_J\,F\,Y = \El\, A \to \Sigma C : \U.\, (\El\, C \to Y) \times F\, C$, 
%we define
%  \begin{eqnarray*}
%    f_0 & \in & \El\, A\to \U \\
%    f_{10} & \in & \Pi a\in \El\, A.\, \El\,(f_0\,a) \to Y\\
%    f_{11} & \in & \Pi a\in \El\, A.\,F\,(f_0\,a)
%  \end{eqnarray*}
%  by $f_0 \eqdf \pi_0 \comp f$, $f_{10} \eqdf \pi_0 \comp \pi_1 \comp f$,
%  $f_{11} \eqdf \pi_1 \comp \pi_1 \comp f$.  
%
  we define
  ${L^F_{A, Y}}^{-1}\,f \in \Lan_J\, (\El\,A \to F\,-)\,Y = 
  \Sigma C : \U.\, (\El\, C \to Y) \times (\El\, A \to F\, Y)$ by
\[ 
{L^F_{A,Y}}^{-1}\,f \eqdf (\sigma\, A\, f_0, 
  \lambda (a,c).\, f_{10}\,a\,c,
  \lambda  a.\, F\,(\lambda c.\, (a,c))\,(f_{11}\, a))
\]
  where $f_0 \in \El\, A\to \U$, $f_{10} \in  \Pi a\in \El\,
  A.\, \El\,(f_0\,a) \to Y$ and $f_{11} \in \Pi a\in \El\,
  A.\, F\,(f_0\,a)$ are defined by $f_0 \eqdf \pi_0 \comp f$, $f_{10}
  \eqdf \pi_0 \comp \pi_1 \comp f$, $f_{11} \eqdf \pi_1 \comp \pi_1
  \comp f$.
 
 We omit the verification that the equivalence relations are
 preserved. We show that ${L^F_{A,Y}}^{-1}$ is indeed the inverse of
 $L^F_{A,Y}$. To prove that it is a left inverse:

   \begin{eqnarray*}
{L^{F}_{A,Y}}^{-1}\,(L^{F}_{A,Y}\,(C,g,k))
     & = & {L^{F}_{A,Y}}^{-1}(\lambda a.\, ( C, g , k\, a)) \\
     & = & (\sigma\, A\, (\lambda a.\, C) , \lambda (a,c).\, g\, c, \lambda a.\, F\,(\lambda c.\, (a,c))\,(k\, a)) \\
     & = & \quad \{ (*) \} \\
     &   & (C , g, k)
     \end{eqnarray*}
To establish (*), we use $\pi_1 \in \El\,(\sigma\, A\, (\lambda a.\, C)) \to \El\, C$, noting that 
     \[(\lambda (a,c).f\, c) = f \comp \pi_1\] and 
     \begin{eqnarray*}
       \lefteqn{(\El\, A \to F \,\pi_1)\,(\lambda a.\, F\,(\lambda c.\, (a,c))\,(k\, a))}\\
       & = & \lambda a.\, (F\,\pi_1 \circ F\,(\lambda c.\, (a,c)))\,(k\, a) \\
       & = & \lambda a.\, F\, \id\,(k\, a) \\
       & = & \lambda a.\, k \, a \\
       & = & k 
     \end{eqnarray*}
     
For the other direction:
       \begin{eqnarray*}
L^{F}_{A,Y}\, ({L^{F}_{A,Y}}^{-1}\,f)
         & = & L^F_{A,Y}\, (\sigma\, A\, f_0, \lambda (a,c).\, f_{10}\,a\,c, \lambda a.\, F\,(\lambda c.\, (a,c))\,(f_{11}\, a))\\
         & = & \lambda a.\, (\sigma\, A\, f_0, \lambda (a,c).\, f_{10}\,a\,c, F\,(\lambda c.\, (a,c))\,(f_{11}\, a))\\
         & = &  \quad \{ (*) \} \\
         &   & \lambda a.\, (f_0\,a, f_{10}\,a, f_{11}\,a) \\
         & = & f
       \end{eqnarray*}
       To justify (*), we exploit, for any $a\in \El\,A$, the function $\lambda c.\, (a, c) \in \El\,(f_0\,a) \to \El\,(\sigma\, A\, f_0)$. %We have 
%\[
%(\lambda (a,c).\, f_{10}\,a\,c) \comp \lambda c.\, (a, c) = f_{10}\,a
%\]

\end{enumerate}
\qed
  
\begin{cor}
  The functor $\Jfin \in \Fin \to \Set$ is well-behaved.
\end{cor}
\proof Choose $\U \eqdf \Nat$ and $\El\, n \eqdf \n$. Clearly this
universe contains $1$ and is closed under $\Sigma$.  \qed

From the well-behavedness of $\Jfin$, it follows that $[\Fin,\Set]$ is
monoidal and $\Lam$ is a monoid. These facts were proved by Fiore et
al.~\cite{FPT:abssvb}.

Our theorem is not general enough to show that $\Jty \in \Fin
\downarrow \Ty \to [\Ty, \Set]$ from Example~\ref{ex:tlam} is
well-behaved, but it ought to be possible to generalize the
construction beyond the case of $\C = \Set$.

\subsection{$[\J,\C]$ is monoidal}

Our well-behavedness conditions suffice to ensure that the unital and
associativity laws of the skew-monoidal structure on $[\J, \C]$ are
isomorphisms. Specifically, the existence of inverses of $J, K, L$
ensures that $\rho, \lambdax, \alphax$ (and consequently also
$\lambda$, $\alpha$) have inverses too.

\begin{thm}\label{thm:properly-monoidal}
  If $J \in \J \to \C$ is well-behaved, then 
% \begin{enumerate}
% \item for any $F \in \J \to \C$, $X \in \obj{\J}$, the map
% $(\rhoinv_{F})_X \in \hom{\C}{\Lan_J\, F\, (J\, X)}{F\, X}$
% defined by $(\rhoinv_{F})_X \eqdf \lcase{\lambda g.\, F\, (J^{-1}\,  g)}$ is an inverse of $\rho_{F,X}$;
% \item for any $X \in \obj{\J}$, the map $\lamxinv_{X}  \in \hom{\C}{X}{\Lan_J\, J\, X}$ 
% defined by $\lamxinv_{X} \eqdf K^{-1}\, \iota_{J,X}$ is an inverse of $\lambdax_{X}$;
% \item for any $F, G \in \J \to \C$, $X \in \obj{\J}$, the map \\
% $(\alphaxinv_{F,G})_X  \in \hom{\C}{\Lan_J\, F\, (\Lan_J\, G\, X)}{\Lan_J\, (\Lan_J\, F \fcomp G)\, X}$
% defined by \\
% $(\alphaxinv_{F,G})_X \eqdf \lcase{\lambda g.\, \lcase{\lambda g.\, \lambda g'.\, 
%    \iota\, g \comp \iota\, g'}\, (L^{-1}\, g)}$ is an inverse of $\alphax_{F,G,X}$.
% \end{enumerate}
% Hence,
the category $([\J,\C],J,\fcompJ,\lambda,\rho,\alpha)$ is monoidal.
\end{thm}

\proof
  To show that this category is monoidal, it suffices to show that $\rho,
  \lambdax, \alphax$ have inverses.

\begin{enumerate}
\item We define, for any $F \in \J \to \C$, $X \in \obj{\J}$, 
\[
\begin{array}{l}
(\rhoinv_{F})_X \in \hom{\C}{\Lan_J\, F\, (J\, X)}{F\, X} \\
(\rhoinv_{F})_X \eqdf \lcase{\lambda g.\, F\, (J^{-1}\,  g)}
\end{array}
\]
We get
\begin{eqnarray*}
(\rhoinv_{F})_X \comp (\rho_{F})_X 
& = & \lcase{\lambda g.\, F\, (J^{-1}\, g)} \comp \iota_{F, J\, X}\, \id_{J\, X} \\
& = & F\, (J^{-1}\, \id_{J\, X}) \\
& = & F\, (J^{-1}\, (J\, \id_X)) \\
& = & F\, \id_X \\
& = & \id_{F\, X}
\end{eqnarray*}
and
\begin{eqnarray*}
(\rho_{F})_X \comp (\rhoinv_{F})_X 
& = & \iota_{F, J\, X}\, \id_{J\, X} \comp \lcase{\lambda g.\, F\, (J^{-1}\, g)} \\
& = & \lcase{\lambda g.\, \iota_{F, J\, X}\, \id_{J\, X} \comp F\, (J^{-1}\, g)} \\
& = & \qquad \mbox{\{~by naturality of $\iota_{F, J\, X}$~\}}\\
&   & \lcase{\lambda g.\, \iota_{F, J\, X}\, (J\, (J^{-1}\, g))}\\
& = & \lcase{\iota_{F, J\, X}} \\
& = & \id_{\Lan_J\, F\ (J\, X)}
\end{eqnarray*}
by the definitions of $\rho_F$, $\rhoinv_F$, the laws of $\Lan_J$ and
$J^{-1}$ being inverse to $J$.

\item We define, for any $F \in \J \to \C$, $X \in \obj{\J}$,
\[
\begin{array}{l}
\lamxinv_X  \in \hom{\C}{X}{\Lan_J\, J\,X} \\
\lamxinv_X \eqdf K^{-1}\, \iota_{J, X}
\end{array}
\]
This gives
\begin{eqnarray*}
\lamxinv_X \comp \lambdax_X 
& = & K^{-1}\, \iota_{J, F\, X} \comp \lcase{\lambda g.\, g} \\
& = & \lcase{\lambda g.\, K^{-1}\, \iota_{J, X} \comp g} \\
& = & \lcase{K\, (K^{-1}\, \iota_{J, X})} \\
& = & \lcase{\iota_{J, X}} \\
& = & \id_{\Lan_J\, J\, X}
\end{eqnarray*}
and
\begin{eqnarray*}
\lambdax_X \comp \lamxinv_X 
& = & \lcase{\lambda g.\, g} \comp K^{-1}\, \iota_{J, X} \\
& = & \qquad \mbox{\{~by naturality of $K^{-1}$~\}} \\
&   & K^{-1}\, (\lambda g.\, \lcase{\lambda g.\, g} \comp \iota_{J, X}\, g)  \\
& = & K^{-1}\, (\lambda g.\, g) \\
& = & K^{-1}\, (K\, \id_{X}) \\
& = & \id_{X}
\end{eqnarray*}
by the definitions of $\lambdax$, $\lamxinv$, $K$, the laws
of $\Lan_J$ and $K^{-1}$ being inverse to $K$.

\item We define, for any $F, G \in \J \to \C$, $X \in \obj{\J}$,
\[
\begin{array}{l}
(\alphaxinv_{F,G})_X  \in \hom{\C}{\Lan_J\, F\, (\Lan_J\, G\, X)}{\Lan_J\, (\Lan_J\, F \cdot G)\, X} \\
(\alphaxinv_{F,G})_X \eqdf \lcase{\lambda g.\, \lcase{\lambda g.\, \lambda g'.\, 
   \iota\, g \comp \iota\, g'}\, (L^{-1}\, g)}
\end{array}
\]
We first observe that 
\begin{eqnarray*}
(\alphax_{F,G})_X 
& = & 
   \lcase{\lambda g.\, \lcase{\lambda g'.\, \iota\, (\iota\, g \comp g')}} \\
& = & 
   \lcase{\lambda g.\, \lcase{\lambda g'.\, \iota\, ((\lambda g'.\, \iota\, g \comp g')\, g')}} \\
& = & 
   \lcase{\lambda g.\, \lcase{\lambda g'.\, \iota\, ((\lcase{\lambda g.\, \lambda g'.\, \iota\, g \comp g'} \comp \iota\, g)\, g')}} \\
& = & 
   \lcase{\lambda g.\, \lcase{\lambda g'.\, \iota\, (\lcase{\lambda g.\, \lambda g'.\, \iota\, g \comp g'}\, (\iota\, g\, g')}} \\
& = & 
   \lcase{\lambda g.\, \lcase{\lambda g'.\, \iota\, (L\, (\iota\, g\, g')}}
\end{eqnarray*}
by the definitions of $\alphax_{F,G}$, $L$ and the laws of $\Lan_J$. 
This observation, together with the definitions of $\alphaxinv_{F,G}$,
the laws of $\Lan_J$ and $L^{-1}$ being inverse to $L$, allows us
to verify
\begin{eqnarray*}
(\alphaxinv_{F,G})_X \comp (\alphax_{F,G})_X 
& = & (\alphaxinv_{F,G})_X
  \comp 
   \lcase{\lambda g.\, \lcase{\lambda g'.\, \iota\, (L\, (\iota\, g\, g')}} \\
& = & 
\lcase{\lambda g.\, (\alphaxinv_{F,G})_X \comp \lcase{\lambda g'.\, \iota\, (L\, (\iota\, g\, g')}} \\
& = & 
\lcase{\lambda g.\, \lcase{\lambda g'.\, (\alphaxinv_{F,G})_X \comp \iota\, (L\, (\iota\, g\, g')}} \\     
& = &
   \lcase{\lambda g.\, \lcase{\lambda g'.\, 
      \lcase{\lambda g.\, \lcase{\lambda g.\, \lambda g'.\, 
      \iota\, g \comp \iota\, g'}\, (L^{-1}\, g)} \comp \iota\, (L\, (\iota\, g\, g')}}  \\
& = &
   \lcase{\lambda g.\, \lcase{\lambda g'.\, 
      \lcase{\lambda g.\, \lambda g'.\, 
      \iota\, g \comp \iota\, g'}\, (L^{-1}\, (L\, (\iota\, g\, g')))}} \\
& = &
   \lcase{\lambda g.\, \lcase{\lambda g'.\, 
      \lcase{\lambda g.\, \lambda g'.\, 
      \iota\, g \comp \iota\, g'}\, (\iota\, g\, g')}} \\
& = &
   \lcase{\lambda g.\, \lcase{\lambda g'.\, 
      (\lcase{\lambda g.\, \lambda g'.\, 
      \iota\, g \comp \iota\, g'} \comp \iota\, g) g'}} \\
& = &
   \lcase{\lambda g.\, \lcase{\lambda g'.\, 
      (\lambda g'.\, \iota\, g \comp \iota\, g') g'}} \\
& = &
   \lcase{\lambda g.\, \lcase{\lambda g'.\, 
      \iota\, g \comp \iota\, g'}} \\
& = &
   \lcase{\lambda g.\, \iota\, g \comp \lcase{\lambda g'.\, \iota\, g'}} \\
& = & 
   \lcase{\lambda g.\, \iota\, g} \\
& = & 
   \id_{\Lan_J\, (\Lan_J\, F \cdot G)\, X}
\end{eqnarray*}
and
\begin{eqnarray*}
(\alphax_{F,G})_X \comp (\alphaxinv_{F,G})_X 
& = & (\alphax_{F,G})_X 
\comp \lcase{\lambda g.\, \lcase{\lambda g.\, \lambda g'.\, 
      \iota\, g \comp \iota\, g'}\, (L^{-1}\, g)} \\
& = & \lcase{\lambda g.\, (\alphax_{F,G})_X \comp \lcase{\lambda g.\, \lambda g'.\, 
      \iota\, g \comp \iota\, g'}\, (L^{-1}\, g)} \\
& = & \qquad \mbox{\{~by definition of $\Y$ (Yoneda embedding)~\}}\\
&   & \lcase{\Y\, (\alphax_{F,G})_X \comp  \lcase{\lambda g.\, \lambda g'.\, 
      \iota\, g \comp \iota\, g'}\, \comp L^{-1} } \\
& = & \lcase{\lcase{\lambda g.\, \Y\, (\alphax_{F,G})_X \comp (\lambda g'.\, 
      \iota\, g \comp \iota\, g')}\, \comp L^{-1} } \\
& = & \qquad \mbox{\{~by definition of $\Y$~\}} \\
&   & \lcase{\lcase{\lambda g.\, \lambda g'.\, (\alphax_{F,G})_X \comp 
      \iota\, g \comp \iota\, g'}\, \comp L^{-1} } \\
& = & \lcase{\lcase{\lambda g.\, \lambda g'.\, 
\lcase{\lambda g.\, \lcase{\lambda g'.\, \iota\, (L\, (\iota\, g\, g')}}
\comp 
      \iota\, g \comp \iota\, g'} \comp L^{-1}} \\
& = & \lcase{\lcase{\lambda g.\, \lambda g'.\, 
\lcase{\lambda g'.\, \iota\, (L\, (\iota\, g\, g')}
\comp \iota\, g'} \comp L^{-1}} \\
& = & \lcase{\lcase{\lambda g.\, \lambda g'.\, 
\iota\, (L\, (\iota\, g\, g')} \comp L^{-1}} \\
& = & \lcase{\lcase{\lambda g.\, 
\iota \comp L \comp \iota\, g} \comp L^{-1}} \\
& = & \lcase{\iota \comp L \comp \lcase{\iota} \comp L^{-1}} \\
& = & \lcase{\iota \comp L \comp L^{-1}} \\
& = & \lcase{\iota} \\
& = & \id_{\Lan_J\, F\, (\Lan_J\, G\, X)}
\end{eqnarray*}\vspace{-30 pt}

\qed
\end{enumerate}

\noindent As an immediate corollary, we get that, in the well-behaved case,
relative monads are proper monoids in a properly monoidal structure.

\begin{cor}
  If $J \in \J \to \C$ is well-behaved, then the category $\RMon{J}$ of
  relative monads on $J$ is isomorphic to the category of monoids in
  the monoidal category $([\J,\C],J,\fcompJ,\lambda,\rho,\alpha)$.
\end{cor}

\subsection{Relative monads extend to monads}
\label{sec:relmon-ext-monad}

As a pleasant bonus, the well-behavedness conditions also ensure that
a relative monad extends to a monad.  Crucial here is that, if $J$ is
well-behaved, then $\lambdax$ and $\alphax$ are isomorphisms. 

% And more, from $\rho$ being an isomorphism too, we get that
% restriction of monads to relative monads undoes extension.

% \begin{align*}
%   & \Tx \in \obj{[\C,\C]} \\
%   & \Tx \eqdf \Lan_J\,T\\
%   & \etax \in \hom{[\C,\C]}{I}{\Tx} \\
% %  & \etax \eqdf \Lan_J\, \eta \comp \lamxinv \\
%   & \etax \eqdf \xymatrix{I \ar[r]^-{\lamxinv} &  \Lan_J\, J \ar[r]^-{\Lan_J\, \eta} & \Lan_J\, T} \\ 
%   & \mux \in \hom{[\C,\C]}{\Tx \fcomp \Tx}{\Tx} \\
% %  & \mux \eqdf \Lan_J \mu \comp \alphayinv_{T,T}
%   & \mux \eqdf \xymatrix{\Lan_J\, T \cdot \Lan_J\, T \ar[r]^-{\alphayinv_{T,T}} & 
%    \Lan_J\, (\Lan_J\, T \cdot T) \ar[r]^-{\Lan_J\, \mu} & \Lan_J\, T}
% \end{align*}

\begin{thm}\label{thm:relmon-ext-mon} 
  Assume that $J \in \J \to \C$ is well-behaved.\\
   A monoid $(T, \eta, \mu)$ in $[\J,\C]$ (equivalently, a relative
  monad on $J$) extends to a monoid $(\Tx, \etax, \mux)$ in $[\C,\C]$
  (equivalently, a monad on $\C$), defined by
\[
\begin{array}{rcl}
% \Tx & \in & \obj{[\C,\C]} \\
\Tx & \eqdf & \Lan_J\, T \\ 
%\etax & \in & \hom{[\C,\C]}{I}{\Tx} \\
\etax & \eqdf
  & \xymatrix{I \ar[r]^-{\lamxinv} & \Lan_J\, J \ar[r]^-{\Lan_J\, \eta}
    & \Lan_J\, T} \\ 
% \mux & \in & \hom{[\C,\C]}{\Tx \fcomp \Tx}{\Tx} \\
\mux & \eqdf 
  & \xymatrix{\Lan_J\, T \cdot \Lan_J\,
    T \ar[r]^-{\alphaxinv_{T,T}} & \Lan_J\, (\Lan_J\, T \cdot T)
    \ar[r]^-{\Lan_J\, \mu} & \Lan_J\, T}
\end{array}
\]
% (2) The correspondence between relative monads on $J$ and monads on
% $\C$ via $(-)^\flat$ and $(-)^\sharp$ is an embedding-projection pair:
% for a relative monad $T$, we have $(T^\sharp)^\flat \cong T$.
\end{thm}

\proof We verify the three monad laws of $T^\sharp$ by the following
diagrams using the respective relative monad laws of $T$, the fact
that $\alphaxinv$ is natural, and one the conditions (b'), (c') and (e') in
each case.

\[
\xymatrix{
\Lan_J\, T \ar[rr]_-{\Lan_J\, T \fcomp \lamxinv} \ar@{=}[ddrr]
   \ar@/^2pc/[rrrr]^{T^\sharp \fcomp \eta^\sharp}
& & \Lan_J\, T \fcomp \Lan_J\, \eta  \ar[rr]_{\Lan_J\, T \fcomp \Lan_J\, \eta}
       \ar[d]^{\alphaxinv_{T,J}}
    & & \Lan_J\, T \fcomp \Lan_J\, T \ar[d]_{\alphaxinv_{T,T}}
                 \ar@/^3.7pc/[ddd]^{\mu^\sharp} \\
& & \Lan_J\, (\Lan_J\, T \fcomp J) \ar[rr]^{\Lan_J\, (\Lan_J\, T \fcomp \eta)} 
    & & \Lan_J\, (\Lan_J\, T \fcomp T) \ar[dd]_{\Lan_J\, \mu} \\
& & \Lan_J\, T \ar[u]_{\Lan_J\, \rho_T} \ar@{=}[drr] 
    & & \\
& & & & \Lan_J\, T
}
\]
\[
\xymatrix{
\Lan_J\, T \ar[dd]^{\lamxinv \fcomp \Lan_J\, T} \ar@{=}[rrdd] 
     \ar@/_3.7pc/[dddd]_{\eta^\sharp \fcomp T^\sharp} \\
\\
\Lan_J\, T \fcomp \Lan_J\, T \ar[dd]^{\Lan_J\, \eta \fcomp \Lan_J\, T}
     \ar[r]_{\alphaxinv_{J,T}}
& \Lan_J\, (\Lan_J\, J \fcomp T) \ar[r]_-{\Lan_J\, \lambda_T}
     \ar[dd]^{\Lan_J\, (\eta \fcomp \Lan_J\, T)} 
  & \Lan_J\, T \ar@{=}[ddr] \\
\\
\Lan_J\, T \fcomp \Lan_J\, T \ar[r]^{\alphaxinv_{T,T}}
   \ar@/_2pc/[rrr]_{\mu^\sharp}
& \Lan_J\, (\Lan_J\, T \fcomp T) \ar[rr]^{\Lan_J\, \mu}
    & & \Lan_J\, T
}    
\]
\[
\scriptsize
\xymatrix@C=0.05pc{
\Lan_J\, T \fcomp \Lan_J\, T \fcomp \Lan_J\, T 
   \ar[rrr]_{\Lan_J\, T \fcomp \alphaxinv_{T,T}}
   \ar[ddd]^{\alphaxinv_{T,T} \fcomp \Lan_J\, T}
   \ar@/^2pc/[rrrrr]^{T^\sharp \fcomp \mu^\sharp}
   \ar@/_5pc/[ddddd]^{\mu^\sharp \fcomp T^\sharp}
& & & \Lan_J\, T \fcomp \Lan_J\, (\Lan_J\, T \fcomp T)
       \ar[rr]_*+{\labelstyle \Lan_J\, T \fcomp \Lan_J\, \mu}
       \ar[dd]_{\alphaxinv_{T,\Lan_J\, T \fcomp T}}
      & & \Lan_J\, T \fcomp \Lan_J\, T 
       \ar[dd]_{\alphaxinv_{T,T}} 
       \ar@/^3.5pc/[ddddd]_{\mu^\sharp}\\
\\
& & & \Lan_J\, (\Lan_J\ T \fcomp \Lan_J\, T \fcomp T)
       \ar[rr]_*+{\labelstyle \Lan_J\, (\Lan_J\, T \fcomp \mu)}
      & & \Lan_J\, (\Lan_J\, T \fcomp T) 
       \ar[ddd]_{\Lan_J\, \mu} \\
\Lan_J\, (\Lan_J\, T \fcomp T) \fcomp \Lan_J\, T
   \ar[rr]^{\alphaxinv_{\Lan_J\, T \fcomp T, T}}
   \ar[dd]^{\Lan_J\, \mu \fcomp \Lan_J\, T}
& & \Lan_J\, (\Lan_J\, (\Lan_J\, T \fcomp T) \fcomp T)
   \ar[ur]^(0.35)*+{\labelstyle \Lan_J\, \alpha_{T,T,T}}
   \ar[dd]^{\Lan_J\, (\Lan_J\, \mu \fcomp T)} \\
\\
\Lan_J\, T \fcomp \Lan_J\, T \ar[rr]^{\alphaxinv_{T,T}}
    \ar@/_2pc/[rrrrr]_{\mu^\sharp}
& & \Lan_J\, (\Lan_J\, T \fcomp T) \ar[rrr]^{\Lan_J\, \mu}
    & & & \Lan_J\, T
}
\]
\qed

Similarly, relative monad morphisms extend to monad morphisms. 

\begin{thm}
  Assume that $J \in \J \to \C$ is well-behaved.
\begin{enumerate}
\item A morphism $\sigma$ between relative monads $T$ and $T'$ on $J$
  extends to a morphism $\sigma^\sharp$ between monads $T^\sharp$ and
  $T'^\sharp$ on $\C$ via $\sigma^\sharp \eqdf \Lan_J\, \sigma$.
\item $(-)^\sharp$ is functorial.
\end{enumerate}
\end{thm}
\newpage

\proof\hfill
\begin{enumerate}
\item The monad morphism laws of $\sigma^\sharp$ are verified by
the following diagrams from the relative monad morphism laws of
$\sigma$ and naturality of $\alphaxinv$.
\[
\xymatrix{
& & & \Lan_J\, T \ar[dd]_{\Lan_J\, \sigma} \ar@/^1pc/[dd]^{\sigma^\sharp} \\
I \ar[r]^-{\lamxinv} 
    \ar@/^1.2pc/[urrr]^{\eta^\sharp}  \ar@/_1.2pc/[drrr]_{\eta'^\sharp}
  & \Lan_J\, J \ar[urr]^{\Lan_J\, \eta} \ar[drr]_{\Lan_J\, \eta'} \\
& & & \Lan_J\, T'
}
\]
\[
\xymatrix{
\Lan_J\, T \fcomp \Lan_J\, T \ar[r]^{\alphaxinv_{T,T}} 
     \ar[dd]^{\Lan_J\, \sigma \fcomp \Lan_J\, \sigma} \ar@/_1.2pc/[dd]_{\sigma^\sharp \fcomp \sigma^\sharp}
     \ar@/^1.8pc/[rrr]^{\mu^\sharp}
  & \Lan_J\, (\Lan_J\, T \fcomp T) \ar[rr]^-{\Lan_J\, \mu}
      \ar[dd]^{\Lan_J\, (\Lan_J\, \sigma \fcomp \sigma)}
    & & \Lan_J\, T \ar[dd]_{\Lan_J\, \sigma} \ar@/^1.2pc/[dd]^{\sigma^\sharp} \\
\\
\Lan_J\, T' \fcomp \Lan_J\, T' \ar[r]_{\alphaxinv_{T',T'}} 
    \ar@/_1.8pc/[rrr]_{\mu'^\sharp}
  & \Lan_J\, (\Lan_J\, T' \fcomp T') \ar[rr]_-{\Lan_J\, \mu'}
    & & \Lan_J\, T' \\
}
\]

\item Functoriality of $(-)^\sharp$ is immediate from functoriality of
$\Lan_J \in [\J, \C] \to [\C, \C]$, as $\RMon{\J}$ and $\Mon{\C}$
inherit their identities and composition from the corresponding
functor categories $[\J, \C]$ and $[\C, \C]$.  
\qed
\end{enumerate}

\noindent We have learned that, in the well-behaved case, not only do monads
restrict to relative monads (by $(-)^\flat$), but relative monads
extend to monads (by $(-)^\sharp$). This relationship turns out to be
an adjunction: $(-)^\sharp$ is left adjoint to $(-)^\flat$.
Furthermore, the adjunction is a coreflection, i.e., the unit is an isomorphism. 

\begin{thm}\label{thm:adjSharpFlat}
  Assume that $J \in \J \to \C$ is well-behaved. Then $(-)^\sharp$ and
  $(-)^\flat$ form an adjunction between $\RMon{J}$ and $\Mon{\C}$.
  Moreover, this adjunction is a coreflection.
\end{thm}

\proof $\Lan_J \in [\J,\C] \to [\C,\C]$ is left adjoint to $(-) \fcomp
J \in [\C,\C] \to [\J,\C]$ with $\rho_T \in T \to \Lan_J\, T \fcomp
J$ (which is an isomorphism) as the unit on $T$ and
\[
\xymatrix{
\Lan_J\, (T \fcomp J) \ar[r]^-{\alphay_{T,J}}
  & T \fcomp \Lan_J\, J \ar[r]^-{T \fcomp \lambdax}
    & T 
}
\]
as the counit.

Since the identities and composition of $\RMon{J}$ and $\Mon{\C}$ are
those of the functor categories $[\J,\C]$ and $[\C,\C]$, we only need
to verify the unit and counit are a relative monad morphism and a
monad morphism, respectively.

The relative monad morphism laws of $\rho_T$ for a relative monad $T$
are verified by the following diagrams from naturality of $\rho$ and
the properties (a), (b), (d) from Theorem \ref{thm:lax}. % CHECK!!
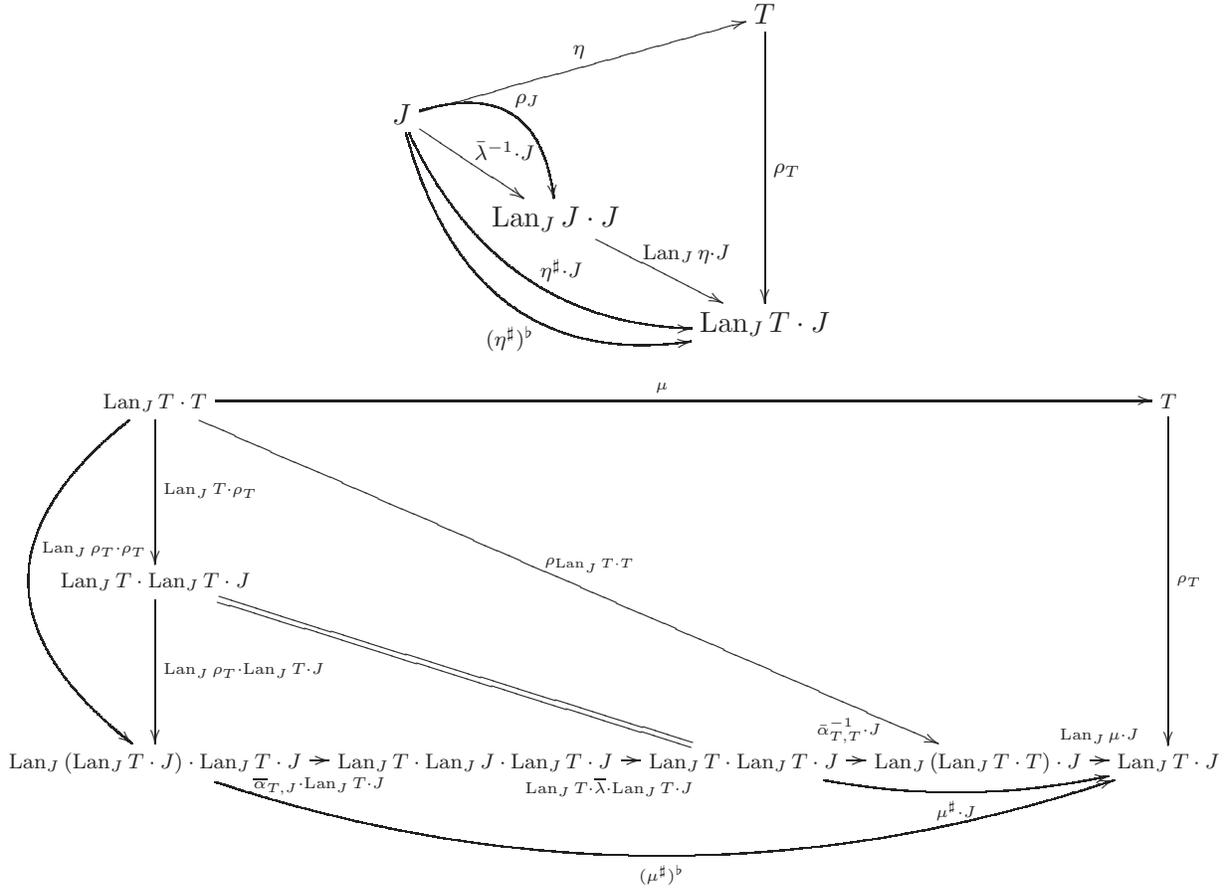
\[
\xymatrix{
& & T \ar[ddd]^{\rho_T} \\
J \ar[rru]^{\eta} \ar@/^2pc/[rd]^{\rho_J} \ar[rd]^{\lamxinv \fcomp J} 
  \ar@/_2.5pc/[rrdd]^{\eta^\sharp \fcomp J}
  \ar@/_3.5pc/[rrdd]_{(\eta^\sharp)^\flat}
  & \\
& \Lan_J\, J \fcomp J \ar[rd]^{\Lan_J\, \eta \fcomp J}
  & \\
& & \Lan_J\, T \fcomp J
}
\]
\[
\hspace*{-5mm}
\scriptsize
\xymatrix@C=0.6pc{
\Lan_J\, T \fcomp T \ar[rrrr]^{\mu}
  \ar[dd]^{\Lan_J\, T \fcomp \rho_T} 
  \ar@/_4pc/[dddd]^(0.4){\Lan_J\, \rho_T \fcomp \rho_T}
  \ar[ddddrrr]^{\rho_{\Lan_J\, T \fcomp T}}
& & & & T \ar[dddd]^{\rho_T} \\
\\
\Lan_J\, T \fcomp \Lan_J\, T \fcomp J\ar@{=}[ddrr]
  \ar[dd]^{\Lan_J\, \rho_T \fcomp \Lan_J\, T \fcomp J} 
\\
\\
\Lan_J\, (\Lan_J\, T \fcomp J) \fcomp \Lan_J\, T \fcomp J
  \ar[r]_-*+{\labelstyle \alphax_{T,J} \fcomp \Lan_J\, T \fcomp J}
  \ar@/_3pc/[rrrr]_{(\mu^\sharp)^\flat}
& \Lan_J\, T \fcomp \Lan_J\, J \fcomp \Lan_J\, T \fcomp J %\Lan_J\, T \fcomp \Lan_J\, J \fcomp \Lan_J\, T \fcomp J
  \ar[r]_*+{\labelstyle \Lan_J\, T \fcomp \lambdax \fcomp \Lan_J\, T \fcomp J}
  & \Lan_J\, T \fcomp \Lan_J\, T \fcomp J
  \ar[r]^(0.45)*+{\labelstyle \alphaxinv_{T,T} \fcomp J}
  \ar@/_1pc/[rr]_{\mu^\sharp \fcomp J}
    & \Lan_J\, (\Lan_J\, T \fcomp T) \fcomp J \ar[r]^-*+{\labelstyle \Lan_J\, \mu \fcomp J}
      & \Lan_J\, T \fcomp J
}
\]

The monad morphism laws of $(T \fcomp \lambdax) \comp \alphay_{T,J}$
for a monad $T$ are verified from naturality of $\eta$ resp.\ $\mu$,
naturality of $\lambdax$ and $\alphay$, and from two elementary
properties of $\alphay$, namely that $\alphay_{I,H} = \Lan_J\, H$ and
$\alphay_{F \fcomp G,H} = (\Lan_J\, F \fcomp \alphay_{G,H}) \comp
\alphay_{F, G \fcomp H}$.
\[
\xymatrix{
& & \Lan_J\, J \ar@/^1pc/[r]^(0.3){\Lan_J\, \eta^\flat} \ar[r]_-*+{\labelstyle \Lan_J\, (\eta \fcomp J)}
    \ar[dl]^(0.3){\alphay_{I,J}}
    & \Lan_J\, (T \fcomp J) \ar[d]^{\alphay_{T,J}}\\
& \Lan_J\, J \ar@/^1pc/@{=}[ur]
    \ar[rr]^{\eta \fcomp \Lan_J\, J}
    \ar[ld]^-(0.3){\lambdax}
  & & T \fcomp \Lan_J\, J \ar[d]^{T \fcomp \lambdax} \\
I \ar@/^1pc/[ru]^(0.65){\lamxinv} \ar@/^7pc/[uurrr]^{(\eta^\flat)^\sharp}
   \ar[rrr]_{\eta}
& & & T 
}
\]
\[
\hspace*{-2mm}
\scriptsize
\xymatrix@C=0.7pc{
\Lan_J\, (T \fcomp J) \fcomp \Lan_J\, (T \fcomp J) 
   \ar@/^3.5pc/[rrrr]^{(\mu^\flat)^\sharp}
   \ar[r]^*+{\labelstyle \alphaxinv_{T \fcomp J, T \fcomp J}} 
   \ar[d]^(0.6){\labelstyle \alphay_{T,J} \fcomp \Lan_J\, (T \fcomp J)}
   \ar[dr]^(0.6)*+{\labelstyle\Lan_J\, (T \fcomp J) \fcomp \alphay_{T,J}}
& \Lan_J\, (\Lan_J\, (T \fcomp J) \fcomp T \fcomp J)
   \ar@/^2pc/[rrr]^{\Lan_J\, \mu^\flat}
   \ar[r]_*+{\labelstyle\Lan_J\, (\alphay_{T,J} \fcomp T \fcomp J)}
   \ar[d]^{\alphay_{\Lan_J\, (T\fcomp J) \fcomp T, J}}
  & \Lan_J\, (T \fcomp \Lan_J \fcomp T \fcomp J)
   \ar[r]_*+{\labelstyle \Lan_J\, (T \fcomp \lambdax \fcomp T \fcomp J)} 
    & \Lan_J\, (T \fcomp T \fcomp J) \ar[r]_*+{\labelstyle \Lan_J\, (\mu \fcomp J)}
%   \ar[ddl]^{\alphay_{T \fcomp \Lan_J\, J \fcomp T, J}}
       & \Lan_J\, (T \fcomp J) 
          \ar[ddd]^{\alphay_{T,J}}\\
T \fcomp \Lan_J\, J \fcomp \Lan_J\, (T \fcomp J)
   \ar[d]^{T \fcomp \lambdax \fcomp \Lan_J\, (T \fcomp J)}
& \Lan_J\, (T\fcomp J) \fcomp T \fcomp \Lan_J\, J
   \ar[d]^{\alphay_{T,J} \fcomp T \fcomp \Lan_J\, J}
%   \ar@/^3pc/[dddlll]^-{\alphay_{T \fcomp T, J}}
  \\
T \fcomp \Lan_J\, (T \fcomp J)
   \ar[d]^{T \fcomp \alphay_{T,J}}
& T \fcomp \Lan_J\, J \fcomp T \fcomp \Lan_J\, J
   \ar[dl]^(0.4)*+{\labelstyle T \fcomp \lambdax \fcomp T \fcomp \Lan_J\, J} \\
T \fcomp T \fcomp \Lan_J\, J
  \ar[d]^{T \fcomp T \fcomp \lambdax}
  \ar[rrrr]^{\mu \fcomp \Lan_J\, J} 
& & & & T \fcomp \Lan_J\, J 
    \ar[d]^{T \fcomp \lambdax} \\
T \fcomp T
   \ar[rrrr]^{\mu}
& & & & T  
}
\]
\qed

We see that, once the extension of relative monads to monads is definable
(which takes that $J$ is well-behaved), it has very good properties
and this happens because the adjunction $\Lan_J \dashv {-} \cdot J$
between $[\J,\C]$ and $[\C,\C]$---the defining adjunction of
$\Lan_J$--- then lifts from functors to (relative) monads. 

Unlike the unit, the counit of this adjunction is generally not an
isomorphism, so the adjunction is not a reflection.  For example, for
$\C \eqdf \Set$, $\J \eqdf \Fin$, $J \eqdf \Jfin$, the $T$-component
of the counit is an isomorphism if and only if the monad $T$ is
finitary. This is important for us: the categories of monads on $\C$
and relative monads on $J$ are generally not equivalent.

\begin{exa}
  For the powerset monad $\P$ on $\Set$, we have that $\P\, X$ is the
  powerset of a set $X$, $\P^\flat\, X \eqdf \P\, (\Jfin\, X)$ is the
  powerset of a finite set $X$, and $\P^{\flat\#}\, X \eqdf
  \Lan_{\Jfin}\, \P^\#\, X$ is the finitary powerset (the set of
  finite subsets) of a (possibly infinite) set $X$. The difference
  between $\P$ and $\P^{\flat\#}$ arises because $\P$ is not finitary.
\end{exa}

\begin{exa}
  For the relative monad $\Vec$ on $\Jfin$, $\Vec^\#\, X$ is the space
  of vectors over a possibly infinite coordinate system $X$ that may
  only have finitely many non-zero components.
\end{exa}

\begin{exa}
  For the relative monad $\Lam$ on $\Jfin$, we have that $\Lam\, X$ is
  the set of $\lambda$-terms over a finite, nameless context $X$ and
  $\Lam^\#\, X$ is given by the set of $\lambda$-terms over a possibly
  infinite, name-carrying context $X$.  The functor $\Lam^\#$ is the
  carrier of the initial algebra of the functor $F \in [\Set,\Set] \to
  [\Set,\Set]$ defined by $F\, G\, X \eqdf X + G\, X \times G\, X +
  G\, (1+X)$.

  For the relative monad $\Lam^\infty$ the picture is different.
  $\Lam^\infty\, X$ is the set of non-wellfounded $\lambda$-terms over a
  finite, nameless context, but ${\Lam^\infty}^\#\, X$ is the set of
  non-wellfounded $\lambda$-terms using a finite number of variables from
  a possibly infinite, name-carrying context. This differs from the
  non-finitary carrier of the final coalgebra of $F$, capturing
  general non-wellfounded $\lambda$-terms that may use infinitely many
  variables.
\end{exa}

The special case where the $T$-component of the counit of $(-)^\#
\dashv (-)^\flat$ is an isomorphism (i.e., $(T^\flat)^\sharp \cong T$)
corresponds to the notion of monad with arities of Berger et
al.~\cite{arities}. A monad on a category $\C$ with a dense
subcategory $\J$ (included in $\C$ via $J \in \J \to \C$) is a monad
with arities if $(T^\flat)^\sharp \cong T$ and if the nerve functor
$K$ corresponding to $J$ preserves $\Lan_J\, T^\flat$ (see
\cite{grellois}). We can see that Berger et al.\, work under our
well-formedness conditions, except that the third condition is only
required of $T^\flat$. In this situation, the associativity law
$\alpha$ of the skew-monoidal category $[\J,\C]$ need not be an
isomorphism, but the component $\alpha_{T^\flat,T^\flat,T^\flat}$ is.

\subsection{Kleisli and Eilenberg-Moore constructions and  extension}
\label{sec:kleisli-em-sharp}

We now explore the relationship between the Kleisli and
Eilenberg-Moore constructions of a given relative monad $T$ on $J$ and
the monad $T^\sharp$ on $\C$.

We assume that $\Lan_J$ exists, that $J$ is dense and satisfies the
3rd well-behavedness condition (so that $\lambda$ and $\alpha$ have
inverses---only then is $T^\sharp$ defined) and optionally also that
$J$ is fully-faithful (so that $\rho$ has also an inverse and
$(T^\sharp)^\flat \cong T$).

There is a functor $D \in \Kl{T} \to \Kl{T^\sharp}$ defined by
\begin{itemize}
\item for any $X \in |\J|$, $D\, X \eqdf J\, X$,
\item for any $X, Y \in |\J|$, $k \in \C(J\, X, T\, Y)$, 
$D\, k \eqdf \xymatrix{ J\, X  \ar[r]^k &  T\, Y  \ar[r]^-{(\rho_T)_Y}  & \Lan_J\, T\, (J\, Y) }$.
\end{itemize}
To prove that $D$ preserves the identities and composition of
$\Kl{T}$, the laws of the monoidal structure on $[\J,\C]$ must be
invoked.

Let $L, R$ be the Kleisli relative adjunction of $T$, which is given
by $L\, X \eqdf X$, $L\, f \eqdf \eta \comp J\, f$, $R\, X \eqdf T\,
X$, $R\, k \eqdf k^*$.

The Kleisli adjunction of $T^\sharp$ is given by $L' X \eqdf X$, $L' f
\eqdf \eta^\sharp \comp f \eqdf \Lan_J\, \eta \comp \lamxinv \comp f$,
$R'\, X \eqdf T^\sharp X = \Lan_J\, T\, X$, 
$R'\, k \eqdf k^{(*^\sharp)} = \mu^\sharp \comp T^\sharp\, k 
= \Lan_J\, \mu \comp \alphaxinv_{T,T} \comp \Lan_J\, T\, k$

We have $D \fcomp L = L' \fcomp J$ and $R = R' \fcomp D$.
As soon as $J \in \J \to \C$ is fully faithful (so that $\rho$ also
has an inverse), $D$ (whose action on objects is $J$) is fully
faithful too.
Moreover, under the same condition, $T$ splits through $\Kl{T^\sharp}$
via $L' \cdot J$ and $R'$: we have $R' \fcomp (L' \fcomp J) = \Lan_J\, T
\cdot J \cong T$ and $L' \cdot J$ is relative left adjoint to $R'$.

No functor is generally definable in the opposite direction
$\Kl{T^\sharp} \to \Kl{T}$.

There is a functor $E \in \EM{T^\sharp} \to \EM{T}$, given by
\begin{itemize}
\item for any $(X, x) \in \EM{T^\sharp}$, i.e., $X \in |\C|$, $x \in
  \C(\Lan_J T X, X)$, subject to EM-algebra conditions, $E\, (X, x)
  \eqdf (X, \chi)$ where, for $Z \in |\J|$, $g \in \C(J\, Z, X)$,
  $\chi_Z\, g \eqdf x \comp \iota\, g \in \C(T\, Z, X)$; 
  $E\, (X, x)$ is a relative EM-algebra for $T$.
\item for any $h \in \EM{T^\sharp}((X, x), (Y, y))$, 
    which is a map in $\C(X, Y)$ satisfying the EM-algebra map
  conditions, 
  $E\, h \eqdf h$, satisfying the relative EM-algebra map conditions.
\end{itemize}

There is also a functor $E^{-1} \in \EM{T} \to \EM{T^\sharp}$ in the
opposite direction, given by
\begin{itemize}
\item for any $(X, \chi) \in \EM{T}$, i.e.,  $X \in |\C|$, 
    for any $Z \in |\J|$, $\chi \in \C(J\, Z, X) \to \C(T\, Z, X)$, 
    subject to the relative EM-algebra conditions,
  $E^{-1}\, (X, \chi) \eqdf (X, x)$ where
    $x = \lcase{\chi} \in \C(\Lan_J\, T\, X, X)$; 
  $E^{-1}\, (X, \chi)$ is an EM-algebra for $T^\sharp$;
\item for any $h \in \EM{T}((X, x), (Y, y))$, 
    which is a map in $\C(X, Y)$ satisfying the relative EM-algebra map 
     conditions, 
  $E^{-1}\, h \eqdf h$, which satisfies the EM-algebra map conditions.
\end{itemize}

\noindent That the identities and composition are preserved is trivial for both
$E$ and $E^{-1}$.

$E$ and $E^{-1}$ are each other's inverses, i.e., the EM-algebras of
$T^\sharp$ and $T$ are the same thing: %\\
$E^{-1}\, (E\, x) = \lcase{\lambda g.\, x \comp \iota\, g} = x \comp \lcase{\iota} = x$ and 
$E\, (E^{-1}\, \chi) = \lambda g.\, \lcase{\chi} \comp \iota\, g = \lambda
g.\, \chi\, g = \chi$.

We arrive at the following picture:
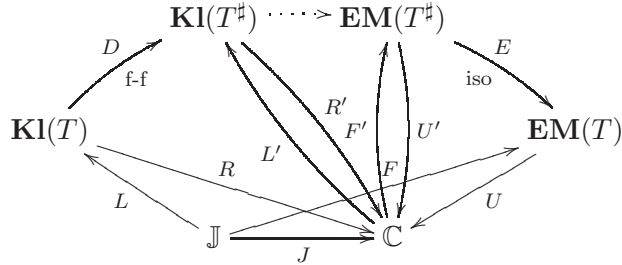
\[
\xymatrix{
& \Kl{T^\sharp} \ar@/^0.2pc/@{.>}[r] \ar@/^0.5pc/[ddr]^{R'}
  & \EM{T^\sharp} \ar@/^0.5pc/[dr]^{E}_{\textrm{iso}} \ar@/^0.5pc/[dd]^{U'}
    &  \\
\Kl{T} \ar@/^0.5pc/[ur]^{D}_{\textrm{f-f}} \ar[drr]^{R}
&  
  & 
    & \EM{T} \ar[dl]^{U} \\ %\ar@/^0.5pc/[ul]^{E^{-1}}\\   
& \J \ar[r]_{J} \ar[ul]^{L} \ar[urr]^{F}
  & \C \ar@/^0.5pc/[uul]^{L'} \ar@/^0.5pc/[uu]^{F'}
    & 
}
\]

% \subsection{Relative adjunctions extend to adjunctions}

% \todo{Keep, extend this section?}

% \begin{thm}\label{thm:remadj-ext-adj}
%   If $J$ is well-behaved, then a relative adjunction $(L, R, \phi)$
%   between $J$ and $\D$ extends to an adjunction $(L^\#, R^\#,
%   \phi^\#)$ between $\C$ and $\D$ where $L^\# \eqdf \Lan_J\,L$,
%   $R^\#\eqdf R$, and the bijective map $\phi^\#$ is defined as
%   follows:
% \[
% \begin{array}{lcl}
% \phi^{\#-1}f&\eqdf&\lcase{\phi^{-1}(\lambda g.\,f\comp g)}\\
% \phi^\# f&\eqdf&K^{-1}(\lambda g.\,\phi(f\comp\Lan_J\,F\,g \circ \rho))
% \end{array}
% \]
% \end{thm}

% =======================================================================
% =======================================================================
% =======================================================================

\section{Arrows as a special case of relative monads}
\label{sec:arrows}

We now turn to a whole class of examples, Hughes'
arrows~\cite{Hug:genma}.  As we shall see, arrows are relative monads
on the Yoneda embedding.  Arrows are commonly perceived as a
generalization of monads. With relative monads, this relationship is
turned upside down!\footnote{Since we compare arrows to monads, not
  strong monads, by arrows we mean arrows without strength in this
  paper. This said, our results scale also to strong arrows, but this
  remains outside the scope of this paper. We have proved this
  elsewhere~\cite{tarmo:cmcs10}. Heunen and Jacobs considered
  strong arrows; their analysis of strength was elaborated by
  Asada \cite{Asada}.}

The rigorous definition of arrows by Heunen and
Jacobs~\cite{HJ:arrlma} is as follows.\footnote{In agreement with the
  previous footnote, this definition does not require $\J$ to be
  symmetric monoidal and an arrow to come with a
  $\mathsf{first}$ operation (strength).}

\begin{defi}
  A ($\Set$-valued) \emph{arrow} on a category $\J$ is given by
\begin{itemize}
\item a function $R \in \obj{\J} \times \obj{\J} \to \obj{\Set}$, 
\item for any $X,
Y \in \obj{\J}$, a function $\pure \in \hom{\J}{X}{Y} \to R\, (X, Y)$, 
\item for any $X, Y, Z \in \obj{\J}$, a function ${\rcomp} \in R\,(Y, Z)
\times R\,(X, Y) \to R\,(X, Z)$,
\end{itemize}
satisfying the conditions
\begin{itemize}
\item $\pure\, (g \comp f) = \pure\, g \comp \pure\, f$,
\item $s \rcomp \pure\, \id = s$,
\item $\pure\, \id \rcomp r = r$,
\item $t \rcomp (s \rcomp r) = (t \rcomp s) \rcomp r$.
\end{itemize}
\end{defi}
It follows from the conditions that $R$ is functorial (contravariantly
in the first argument), i.e., $R \in \J^\op \times \J \to \Set$, which
is the same as to say that $R$ is an endoprofunctor on $\J$, and
$\pure$ and $\rcomp$ are natural/dinatural.

A monad $(T, \eta, (-)^*)$ on $\J$ defines an arrow $(R, \pure,
\rcomp)$ on $\J$ by $R\, (X, Y) \eqdf \hom{\Kl{T}}{X}{Y}$, $\pure\, f
\eqdf L\, f$ and $\ell \rcomp k \eqdf \ell \compKl k$ where $L$ is the
left adjoint in the Kleisli adjunction and $\compKl$ is the Kleisli
composition.

We show now that an arrow on $\J$ is the same thing as a relative
monad on the Yoneda embedding $\Y \in \J \to [\J^\op, \Set]$ defined
by $\Y\, X\, Y \eqdf \hom{\J}{Y}{X}$. 

By definition, a relative monad
on $\Y$ is given by
\begin{itemize}
\item a function $T \in \obj{\J} \to \obj{[\J^\op, \Set]}$,
\item for any $X \in \obj{\J}$, a map $\eta_X \in \hom{[\J^\op,
    \Set]}{\Y\, X}{T\, X}$,
\item for any $X, Y \in \obj{\J}$, a map function $(-)^* \in
  \hom{[\J^\op, \Set]}{\Y\, X}{T\, Y} \to \hom{[\J^\op, \Set]}{T\,
    X}{T\, Y}$
\end{itemize}
satisfying three coherence conditions.

\begin{thm}\label{thm:arrow-relmon}
\hfill
\begin{enumerate}
\item An arrow $(R, \pure, \rcomp)$ on $\J$ gives rise to a relative
  monad $(T, \eta, (-)^*)$ on $\Y$ defined by
%\[
$T\, X\, Y \eqdf R\, (Y, X)$,
$T\, \anon\, f\, r \eqdf r \rcomp \pure\, f$,
$\eta\, f \eqdf \pure\, f$,
$k^*\, r \eqdf k\, \id \rcomp r$.
%\]
\item A relative monad $(T, \eta, (-)^*)$ on $\Y$ gives rise to an
  arrow
  $(R, \pure, \rcomp)$ on $\J$ defined by
%\[
$R\, (X, Y) \eqdf T\, Y\, X$,
$\pure\, f \eqdf \eta\, f$,
$s \rcomp r \eqdf (\lambda f.\, T\, \anon\, f\ s)^* r$.
%]
(The last item is well-defined, as $\lambda f.\, T\, \anon\, f\ s$ is
natural.)
\item The above is a bijective correspondence. 
\end{enumerate}
\end{thm}

\proof\hfill
\begin{enumerate}
\item We have to verify functoriality of $T$ and naturality of
$\eta$, $(-)^*$ in their contravariant arguments and the three
relative monad laws. The proofs are as follows.

Proofs of contravariant functoriality of $T$:
\begin{eqnarray*}
T\, \_\, \id\, r 
& = & r \rcomp \pure\, \id  \\
& = & \qquad \mbox{\{~by 2nd arrow law~\}} \\
&   & r 
\end{eqnarray*}
\begin{eqnarray*}
T\, \_\, (g \comp f)\, r 
& = & r \rcomp \pure\, (g \comp f)  \\
& = & \qquad \mbox{\{~by 1st arrow law~\}} \\
&   & r \rcomp (\pure\, g \rcomp \pure\, f) \\
& = & \qquad \mbox{\{~by 4th arrow law~\}} \\
&   & (r \rcomp \pure\, g) \rcomp \pure\, f \\
& = & T\, \_\, f\, (r \rcomp \pure\, g) \\
& = & T\, \_\, f\, (T\, \_\, g\, r) \\
& = & (T\, \_\, f\, \comp T\, \_\, g)\, r 
\end{eqnarray*}

Proofs of contravariant naturality of $\eta$ and $(-)^*$:
\begin{eqnarray*}
\eta\, (g \comp f) 
& = & \pure\, (g \comp f) \\
& = & \qquad \mbox{\{~by 1st arrow law~\}} \\
&   & \pure\, g \rcomp \pure\, f \\
& = & T\, \_\, f\, (\eta\, g) \\
\end{eqnarray*}
\begin{eqnarray*}
k^*\, (T\, \_\, f\, r) 
& = & k\, \id \rcomp (r \rcomp \pure\, f) \\
& = & \qquad \mbox{\{~by 4th arrow law~\}} \\
&   & (k\, \id \rcomp r) \rcomp \pure\, f \\
& = & k^*\, r \rcomp \pure f \\
& = & T\, \_\, f\, (k^*\, r) 
\end{eqnarray*}

Proofs of relative monad laws:
\begin{eqnarray*}
(k^* \comp \eta)\, f 
& = & k^* (\eta\, f) \\
& = & k\, \id \rcomp \eta\, f \\
& = & k\, \id \rcomp \pure\, f \\
& = & T\, \_\, f\, (k\, \id)  \\
& = & \qquad \mbox{\{~by contravar.\ naturality of $k$~\}} \\
&   & k\, (\id \comp f) \\
& = & k\, f
\end{eqnarray*}
\begin{eqnarray*}
\eta^*\, r 
& = & \pure^*\, r \\
& = & \pure\, \id \rcomp r \\
& = & \qquad \mbox{\{~by 3rd arrow law~\}} \\
&   & r  
\end{eqnarray*}
\begin{eqnarray*}
(\ell^* \comp k)^*\, r 
& = & (\ell^* \comp k)\, \id \rcomp r \\ 
& = & (\ell^*\, (k\, \id)) \rcomp r \\
& = & (\ell\, \id \rcomp k\, \id) \rcomp r \\
& = & \quad \mbox{\{~by 4th arrow law~\}} \\
&   & \ell\, \id \rcomp (k\, \id \rcomp r) \\
& = & \ell^*\, (k\, \id \rcomp r) \\
& = & \ell^*\, (k^*\,  r) \\
& = & (\ell^* \comp k^*)\, r
\end{eqnarray*}

\item To see that the definition of $\rcomp$ is wellformed, we must
check that $\lambda f.\, T\, \_\, f\, s$ is natural in the
contravariant argument, which it is.

We can verify all four arrow laws.
\begin{eqnarray*}
\pure\, g \rcomp \pure\, f 
& = & \eta\, g \rcomp \eta\, f \\
& = & (\lambda f'.\, T\, \_\, f'\, (\eta\, g))^* (\eta\, f) \\
&   & \qquad \mbox{\{~by contravar. naturality of $\eta$~\}} \\
& = & (\lambda f'.\, \eta\, (g \comp f'))^* (\eta\, f) \\
& = & ((\lambda f'.\, \eta\, (g \comp f'))^* \comp \eta)\, f \\
& = & \qquad \mbox{\{~by 1st relative monad law~\}} \\
&   & (\lambda f'.\, \eta\, (g \comp f'))\, f \\
& = & \eta\, (g \comp f) \\
& = & \pure\ (g \comp f)
\end{eqnarray*}
\begin{eqnarray*}
r \rcomp \pure\, \id 
& = & (\lambda f.\, T\, \_\, f\, r)^* (\eta\, \id) \\
& = & ((\lambda f.\, T\, \_\, f\, r)^* \comp \eta)\, \id \\
& = & \qquad \mbox{\{~by 1st relative monad law~\}} \\
&   & T\, \_\, \id\, r \\
& = & \qquad \mbox{\{~by contravar.\ functoriality of $T$~\}} \\
&   & r
\end{eqnarray*}
\begin{eqnarray*}
\pure\, \id \rcomp r 
& = & (\lambda f.\, T\, \_\, f\, (\eta\, \id))^*\,  r \\
& = & \qquad \mbox{\{~by contravar.\ naturality of $\eta$~\}} \\
&   &  (\lambda f.\, \eta\, (\id \comp f))^*\, r \\
& = & \eta^*\, r  \\
& = & \qquad \mbox{\{~by 2nd relative monad law~\}} \\
&   & r  
\end{eqnarray*}
\begin{eqnarray*}
(t \rcomp s) \rcomp r 
& = & (\lambda f.\, T\, \_\, f\,  t)^*\, s) \rcomp r \\
& = & (\lambda f.\, T\, \_\, f\, ((\lambda f.\, T\, \_\, f\,  t)^*\, s))^*\, r \\
& = & \qquad \mbox{\{~by contravar.\ naturality of $(-)^*$~\}} \\
&   & (\lambda f.\, (\lambda f.\, T\, \_\, f\,  t)^*\,  (T\, \_\, f\, s))^*\, r \\
& = & (\lambda f.\, (\lambda f.\, T\, \_\, f\,  t)^*\,  ((\lambda f.\, T\, \_\, f\, s)\, f))^*\, r \\
& = & ((\lambda f.\, T\, \_\, f\,  t)^* \comp (\lambda f.\, T\, \_\, f\, s))^*\, r \\
& = & \qquad \mbox{\{~by 3rd relative monad law~\}} \\
& = & ((\lambda f.\, T\, \_\, f\, t)^* \comp (\lambda f.\, T\, \_\, f\, s)^*)\, r \\
& = & (\lambda f.\, T\, \_\, f\, t)^* ((\lambda f.\, T\, \_\, f\, s)^*\, r) \\
& = & t \rcomp (\lambda f.\, T\, \_\, f\, s)^*\, r \\
& = & t \rcomp (s \rcomp r)
\end{eqnarray*}

\noindent The conditions for the bijection (3) just follow from the respective
relative monad and arrow laws except in the case of $k^* r$ where we must use
also invoke the naturality of $k$.\qed\smallskip
\end{enumerate}

\noindent The bijection extends to an isomorphism of the categories of arrows on
$\J$ and relative monads on $\Y$.

\begin{defi}
\label{def:arrow-morph}
A arrow morphism between arrows $(R,\,\pure,\,\rcomp)$ and 
$(R', \pure', \rcomp')$ is given by
\begin{itemize}
\item a function $\tau_{X,Y}\in R\,(X,Y) \to R'\,(X,Y)$
\end{itemize}
satisfying the conditions
\begin{itemize}
\item $\tau\, (\pure\,f) = \pure'\,f$,
\item $\tau\, (f\rcomp g) = \tau\, f \rcomp' \tau\, g$.
\end{itemize} 
\end{defi}
\begin{thm}\hfill

\label{thm:arrow-morph}
\begin{enumerate}
\item An arrow morphism $\tau$ between arrows $(R,\pure,\rcomp)$ and $(R',\pure', \rcomp')$ on $\J$ gives rise to a relative monad morphism $\sigma_{X} \in \hom{[\J^\op,\Set]}{T X}{T' X}$ defined as $\sigma_{X,Y} \eqdf \tau_{Y,X}$ where $T\,X\,Y \eqdf R\,(Y, X)$ and $T'\, X\,Y \eqdf R'\,(Y, X)$.
\item A relative monad morphism $\sigma$ between relative monads $(T,\eta,(-)^*)$ and $(T',\eta',(-)^{*'})$ gives rise to an arrow morphism $\tau$ whose components $\tau_{X,Y} \in R\, (X,Y) \rightarrow R'\,(X,Y)$ are defined as $\tau_{X,Y} \eqdf \sigma_{Y,X}$ where $R\, (X,Y) \eqdf T\, Y\, X$ and $R'\,(X, Y) \eqdf T'\, Y\, X$.
\item The categories of relative monads on $\Y$ and arrows on $\J$ are isomorphic.  
\end{enumerate}
\end{thm}
\proof\hfill
\begin{enumerate}
\item We need to check the relative monad morphism conditions:
\begin{eqnarray*}
\sigma\,(\eta\,f) 
&=& \tau\, (\pure\,f) \\
&=& \qquad\textrm{\{~by pure pres.\, law of $\tau$~\}} \\
& & \pure' f\\ 
&=&  \eta'\,f
\end{eqnarray*}
\begin{eqnarray*}
\sigma\, (k^*\,f)
&=& \tau\, (k\,\id\rcomp f)\\
&=& \qquad\textrm{\{~by compos.\, pres.\, law of $\tau$~\}} \\
& & \tau\,(k\,\id)\rcomp' \tau\, f \\
&=& (\sigma \comp k)^{*'} (\sigma\, f)
\end{eqnarray*}

\item We check the arrow morphism conditions:
\begin{eqnarray*}
\tau\, (\pure\,f) 
&=& \sigma\, (\eta\,f)\\
&=& \qquad\textrm{\{~by unit pres.\ law of $\sigma$~\}} \\
& & \eta'\,f \\ 
&=& \pure'\,f
\end{eqnarray*}
\begin{eqnarray*}
\tau\, (f\rcomp g)
&=& \sigma\, ((\lambda h.\,T\,\_\,h\,f)^*\,g)\\
&=& \qquad\textrm{\{~by Kl. ext.\ pres.\ law of $\sigma$~\}}\\
& & (\sigma\comp(\lambda h.\,T\,\_\,h\,f))^{*'} (\sigma\,g) \\
&=& \qquad\textrm{\{~by naturality of $\sigma$~\}}\\
& & (\lambda h.\,T'\,\_\,h\,(\sigma f))^{*'} (\sigma\,g) \\
&=& \tau\, f \rcomp' \tau\, g
\end{eqnarray*}

\item That the correspondence is an isomorphism is trivial.\qed
\end{enumerate}

It is easy to verify that the Freyd category of an arrow is
the Kleisli category of the corresponding relative monad.
Jacobs et al.~\cite{JHH:catsa} have previously proved that ``Freyd is
Kleisli for arrows'' taking ``Kleisli for arrows'' to mean a
construction that is Kleisli-like under a 2-categorical view of the
Kleisli construction for monads. We can take it to mean ``Kleisli for
arrows as relative monads''.

Similarly to $\Jfin$ and $\JU$ considered above, the functor $\Y$ is
well-behaved. The result of Heunen and Jacobs~\cite{HJ:arrlma} about
arrows being monoids follows as an instance of a generality about
relative monads.

\begin{thm}
The Yoneda embedding $\Y \in \J \to [\J^\op,\Set]$ is well-behaved:
\begin{enumerate}
\item for any $X, Y \in \obj{\J}$, $J_{X,Y}$ has an inverse 
\[
{J_{X,Y}}^{-1} \in \hom{[\J^\op,\Set]}{\Y\,X}{\Y\,Y} \to \hom{\J}{X}{Y}
\]
\item for any $G, H \in \obj{[\J^\op, \Set]}$, $K_{G,H}$ has an inverse 
\[
{K_{G,H}}^{-1} \in \hom{[\J^\op,\Set]}{\hom{[\J^\op,\Set]}{\Y\,-}{G}}{\hom{[\J^\op,\Set]}{\Y\,-}{H}} 
  \to \hom{[\J^\op,\Set]}{G}{H}
\]
\item for any $F \in \J \to [\J^\op,\Set]$, $X \in \obj{\J}$, $H \in \obj{[\J^\op, \Set]}$, $L^F_{X,H}$ has an inverse 
\[
{L^F_{X,H}}^{-1} \in 
\hom{[\J^\op,\Set]}{\Y\,X}{\Lan_\Y\,F\,H} 
\to \Lan_\Y\,(\hom{[\J^\op,\Set]}{\Y\,X}{F\,-})\,H
\]
\end{enumerate} 
\end{thm}

To prove the 3rd item, we use that coends in presheaf categories are
constructed pointwise. We have $\Lan_\Y\,F\,H\, Z \cong \Lan_\Y\,
(F'\, Z)\, H$ where $F'\, Z\, X = F\, X\, Z$.

\proof\hfill
\begin{enumerate}
\item
%\[
%\begin{array}{lcl}
%J & \in & \hom{\J}{A}{B} \rightarrow \int_X\hom{\J}{X}{A} \rightarrow \hom{\J}{X}{B}\\
%J f &\eqdf& \Y f\\
%J^{-1} &\in & (\int_X\hom{\J}{X}{A} \rightarrow \hom{\J}{X}{B}) \rightarrow \hom{\J}{A}{B}\\
%J^{-1} \alpha &\eqdf& \alpha\,\id
%\end{array}
%\]
%\[
%J^{-1}(J\,f)\, = \, J^{-1}(\Y\,f) \, = \, \Y\,f\,\id \, = \, f\,\comp\,\id \, %= \, f
%\]
Recall that $J_{X,Y}\, f \eqdf \Y\, f = \lambda g.\, f \comp g$. By
the Yoneda lemma $J_{X,Y}$ is an isomorphism and the inverse of
$J_{X,Y}$ is
\[
{J_{X,Y}}^{-1}\, \tau \eqdf \tau\, \id
\]

\item The inverse of $K_{G,H}$ is definable by
\[
{K_{G,H}}^{-1} \alpha 
   \eqdf \lambda a.\, \alpha\, (\lambda f.\, G\, f\, a)\, \id 
\]
Proof:
\begin{eqnarray*}
{K_{G,H}}^{-1}\, (K_{G,H}\, \tau) 
& = & {K_{G,H}}^{-1}\, (\lambda \theta.\, \tau \comp \theta) \\
& = & \lambda a.\, (\tau \comp \lambda f.\, G\, f\, a)\, \id \\
& = & \lambda a.\, \tau\, ((\lambda f.\, G\, f\, a)\, \id) \\
& = & \lambda a.\, \tau\, (G\, \id\, a) \\
& = & \lambda a.\, \tau\, a \\
& = & \tau
\end{eqnarray*}

\begin{eqnarray*}
K_{G,H}\, ({K_{G,H}}^{-1}\, \alpha)
& = &  K_{G,H}\, (\lambda a.\, \alpha\, (\lambda f.\, G\, f\, a)\, \id) \\
& = & \lambda \theta.\, (\lambda a.\, \alpha\, (\lambda f.\, G\, f\, a)\, \id) \comp \theta \\
& = & \lambda \theta.\, \lambda g.\, (\lambda a.\, \alpha\, (\lambda f.\, G\, f\, a)\, \id)\, (\theta\, g) \\
& = & \lambda \theta.\, \lambda g.\, \alpha\, (\lambda f.\, G\, f\, (\theta\, g))\, \id \\
& = & \quad \{ \textrm{~by naturality of $\theta$~} \} \\
&   & \lambda \theta.\, \lambda g.\, \alpha\, (\lambda f.\, \theta\, (g \comp f))\, \id \\
& = & \lambda \theta.\, \lambda g.\, \alpha\, (\theta\, \comp (\lambda f.\, g \comp f))\, \id \\
& = & \lambda \theta.\, \lambda g.\, \alpha\, (\theta\, \comp \Y\, g)\, \id \\
& = & \quad \{ \textrm{~by naturality of $\alpha$~} \} \\
&   & \lambda \theta.\, \lambda g.\, (\alpha\, \theta\, \comp \Y\, g)\, \id \\
& = & \lambda \theta.\, \lambda g.\, \alpha\, \theta\, (\Y\, g\, \id) \\
& = & \lambda \theta.\, \lambda g.\, \alpha\, \theta\, (g \comp \id) \\
& = & \lambda \theta.\, \lambda g.\, \alpha\, \theta\, g \\
& = & \alpha
\end{eqnarray*}

In fact, it is immediate to conclude from the Yoneda lemma (applying
it twice) that the sets $\hom{[\J^\op,\Set]}{G}{H}$ and
$\hom{[\J^\op,\Set]}{\hom{[\J^\op,\Set]}{\Y\,-}{G}}{\hom{[\J^\op,\Set]}{\Y\,-}{H}}$
are isomorphic, but we must also check that this isomorphism is
$K_{G,H}$.

%\[
%\begin{array}{lcl}
%K&\in& (\int_Z A\,Z\rightarrow B\,Z) \rightarrow\\
%&& \int_X (\int_Y\hom{\J}{Y}{X} \rightarrow A\,Y)\rightarrow \int_Y\hom{\J}{Y}{X}\rightarrow B\,Y\\
%K\,\alpha\,X\,\beta\,Y\,f &\eqdf& \alpha\,Y\,(\beta\,Y\,f) \\
%K^{-1}&\in& (\int_X (\int_Y \hom{\J}{Y}{X} \rightarrow A\,Y)\rightarrow \int_Y\hom{\J}{Y}{X}\rightarrow B\,Y) \rightarrow\\
%&& \int_Z A\,Z\rightarrow B\,Z\\

%K^{-1}\,\delta\,Z\,a&\eqdf &\delta\,Z\,(\lambda Y,\,h.\,A\,h\,a)\,Z\,\id
%\end{array}
%\]
%\begin{eqnarray*}
%\lefteqn{K\,(K^{-1}\delta)\,X\,\beta\,Y\,f}\\
%& = & K^{-1}\delta\,Y\,(\beta\,Y\,f)\\
%& = & \delta\,\,Y\,(\lambda Z,\,h.\,A\,h\,(\beta\,Y\,f))\,Y\,\id\\
%& = & \delta\,\,Y\,(\lambda Z,\,h.\beta\,Z\,(h\,\comp\,f))\,Y\,\id\quad\{\textrm{by naturality of $\beta$}\}\\
%& = & \delta\,Y\,(\beta\,\comp\,\Y\,X\,f)\,Y\,\id\\
%& = & \delta\,X\,\beta\,Y\,f\quad\{\textrm{by naturality of $\delta$}\}
%\end{eqnarray*}

%\begin{eqnarray*}
%\lefteqn{K^{-1}(K\,\alpha)\,Z\,a}\\
%& = & K\,\alpha\,Z\,(\lambda Y,\,h.\,A\,h\,a)\,Z\,\id\\
%& = & \alpha\,Z\,(A\,\id\,a) \\
%& = & \alpha\,Z\,a
%\end{eqnarray*}

\item That the sets $\Lan_\Y\,(\hom{[\J^\op,\Set]}{\Y\,X}{F\,-})\,H$
  and $\hom{[\J^\op,\Set]}{\Y\,X}{\Lan_\Y\,F\,H}$ are isomorphic
  follows from the Yoneda lemma combined with the fact that
  coends in presheaf categories are constructed pointwise. Again it is
  important to verify that the isomorphism is $L^{F}_{X,H}$.\qed

%\[
%\begin{array}{lcl}
%L &\in& (\int^k (\int_l\hom{\J}{l}{k} \rightarrow G\,l)\times \int_l\hom{\J}{l}{i} \rightarrow F\,k\,l) \rightarrow\\
%&& \int_j\hom{\J}{j}{i} \rightarrow \int^k (\int_l\hom{\J}{l}{k} \rightarrow G\,l) \times F\,k\,j\\
%L\,(k,\,\alpha,\,\beta)& \eqdf& \lambda j,\,f.\,(k,\,\alpha,\,\beta\,j\,f)\\\\
%L^{-1}&\in& (\int_j\hom{\J}{j}{i} \rightarrow \int^k (\int_l\hom{\J}{l}{k} \rightarrow G\,l) \times F\,k\,j) \rightarrow\\
%&&\int^k (\int_l \hom{\J}{l}{k} \rightarrow G\,l)\times \int_l\hom{\J}{l}{i} \rightarrow F\,k\,l\\
%L^{-1} \delta &\eqdf& (k,\,\alpha,\,\lambda\,l,\,g.\,F\,k\,g\,x)\\
%&& \textbf{where}\, (k,\,\alpha,\,x) \eqdf \delta\,\,i\,\id
%\end{array}
%\]
%\begin{eqnarray*}
%\lefteqn{L\,(L^{-1}\delta)\,j\,f}\\
%& = & L (k,\,\alpha,\,\lambda l,\,g.(F\,k\,g\,x))\,j\,f\;\; \textbf{where}\; (k,\,\alpha,\,x) \eqdf \delta\,i\,\id\\
%& = & (k,\,\alpha,\,F\,k\,f\,x)\; \textbf{where}\; (k,\,\alpha,\,x) \eqdf \delta\,i\,\id\\& = & (k,\,\alpha,\,x) \;\textbf{where}\; (k,\,\alpha,\,x) \eqdf \delta\,j\,f\quad\{\textrm{by naturality of $\delta$}\}\\
%& = & \delta\,j\,f
%\end{eqnarray*}

%\begin{eqnarray*}
%\lefteqn{L^{-1}\,(L\,(k,\,\alpha,\,\beta))}\\
%& = & L^{-1}(\lambda j,f.\,k,\,\alpha,\,\beta j f) \\
%& = & (k,\,\alpha,\,\lambda l,g.\,F\,k\,g\, (\beta\,i\, \id))\\
%& = & (k,\,\alpha,\,\lambda l,g.\,\beta\,l\, g)\quad\{\textrm{by naturality of $\beta$}\}\\
%& = & (k,\,\alpha,\,\beta)
%\end{eqnarray*}

\end{enumerate}

\begin{cor}
If $\J$ is small, then, as $\Y$ is well-behaved, the category $[\J,
  [\J^\op, \Set]]$ is monoidal. An arrow on $\J$ is a monoid in this
category.
\end{cor}

Heunen and Jacobs~\cite{HJ:arrlma} considered the special case of
arrows and showed an arrow to be a monoid in $[\J^\op \times \J,
  \Set]$ (the category of endoprofunctors on $\J$) as a monoidal
category, which is, of course, an equivalent statement.

% =======================================================================
% =======================================================================
% =======================================================================

\section{Conclusions and further work}
\label{sec:concl}

We have introduced a generalization of monads, relative monads, which
is motivated by examples and subsumes arrows, a well-known
generalization of monads. Indeed, when moving to a more precise type
discipline, the illusion that everything takes place in only one
ambient category (say, $\Set$) can no longer be maintained and as a
consequence we have to revisit the categorically inspired concepts of
functional programming. We believe that our examples demonstrate that
monad-like entities which are not endofunctors are natural;
fortunately, they are precisely monoids in the functor category.  We
also suggest that our presentation of relative monads given in
Sect.~\ref{sec:relmon} is accessible for functional programmers,
indeed it does not differ substantially from ordinary
monads. 

%Our development is only the first step. Due to lack of space, we have
%not written about monad maps; we did not comment on the relationship
%between relative adjunctions and adjunctions etc.; strong monads
%(esp.\ versus strong arrows) are a further additional topic.
We will elsewhere comment on the relation of our relative monads to
the recent generalization of monads by Spivey \cite{Spi:algcs} that
was also motivated by programming examples: he fixes a functor $K \in
\C \to \J$ (notice the direction) to then look for monad-like
structures with an underlying functor $\J \to \C$. With Paul Levy we
have checked that a fair amount of monad theory transfers to his
generalized monads, but they are not monoids in $[\J,\C]$ unless $K$
has a left adjoint, in which case they are equivalent to relative
monads. Sam Staton has considered an enriched variant of relative
monads~\cite{staton:clones}.

It seems clear that many of the concepts known from ordinary monads
carry over to the relative setting. We hope that this generalization
of the monadic approach leads to new programming structures supporting
a greater reusability of concepts and programs. Indeed, relative
monads have already been used by Ahrens to model syntax with a
reduction
relation~\cite{ahrens:wollic,ahrens:mscs}. Orchard~\cite{orchard} has
generalized monads to relative monads in Haskell using constraint
kinds and associated types. Gabbay and Nanevski \cite{gabbay} needed
relative comonads in their work on contextual modal type theory.

We have formalized a large part of the development of the present
paper in the dependently typed programming language Agda
\cite{thorsten.tarmo.james:jfr}.

% While we were writing this paper we became aware of Berger, Melli\`es
% and Weber's brand new and yet unpublished work on ``monads with
% arities'' \cite{arities}. We have not yet had an opportunity to
% explore this work in any detail, but it it seems that monads with
% arities are relative monads on well-behaved functors.

Skew-monoidal categories are interesting in their own right. We have
recently~\cite{tarmo:coh} proved a coherence theorem for
them---identified a sufficient condition for a unique ``formal'' map
between two given ``formal'' objects. Lack and Street~\cite{LS13}
proved a different one, which is a necessary and sufficient condition
for equality of two given maps.

%We know of some coherence theorems for structures like these
%\cite{Laplaza,Dosen}, but of no work that would cover our case.
% CHECK HOW TO CITE

\paragraph{\bf Acknowledgements}

We are grateful to Paul Levy and Thomas Streicher for valuable
comments and hints, and to the anonymous referees of both this paper
and the conference version on which it is based.
% and to
%several people on the Categories mailing list for answers about
%related or not so related work.

%\bibliographystyle{abbrv}
%\bibliography{lmcs-final}

\end{document}